\documentclass[]{aastex6}
\shorttitle{Angular Momentum of Galaxies}
\shortauthors{Shi et al.}
\usepackage{bm}
\usepackage{natbib}
\begin{document}
\title{Angular Momentum of Early and Late Type Galaxies:\\ Nature or Nurture?}
\author{J. Shi\altaffilmark{1,2}, A. Lapi\altaffilmark{1,3,4}, C. Mancuso\altaffilmark{5}, H. Wang\altaffilmark{2}, L. Danese\altaffilmark{1,3,4}}
\altaffiltext{1}{SISSA, Via Bonomea 265, 34136 Trieste, Italy}
\altaffiltext{2}{Dept. of Astronomy, Univ. of Science and Technology of China, Hefei, 230026 Anhui, China} \altaffiltext{3}{INAF-Osservatorio Astronomico di Trieste, via Tiepolo 11, 34131 Trieste, Italy} \altaffiltext{4}{INFN-Sezione di Trieste, via Valerio 2, 34127 Trieste, Italy}\altaffiltext{5}{INAF-IRA, Via P. Gobetti 101, 40129 Bologna, Italy}

\begin{abstract}
We investigate the origin, the shape, the scatter, and the cosmic evolution in the observed relationship between specific angular momentum $j_\star$ and the stellar mass $M_\star$ in early-type (ETGs) and late-type galaxies (LTGs). Specifically, we exploit the observed star-formation efficiency and chemical abundance to infer the fraction $f_{\rm inf}$ of baryons that infall toward the central regions of galaxies where star formation can occur. We find $f_{\rm inf}\approx 1$ for LTGs and $\approx 0.4$ for ETGs with an uncertainty of about $0.25$ dex, consistent with a biased collapse. By comparing with the locally observed $j_\star$ vs. $M_\star$ relations for LTGs and ETGs we estimate the fraction $f_j$ of the initial specific angular momentum associated to the infalling gas that is retained in the stellar component: for LTGs we find $f_j\approx 1.11^{+0.75}_{-0.44}$, in line with the classic disc formation picture; for ETGs we infer $f_j\approx 0.64^{+0.20}_{-0.16}$, that can be traced back to a $z\la 1$ evolution via dry mergers. We also show that the observed scatter in the $j_{\star}$ vs. $M_{\star}$ relation for both galaxy types is mainly contributed by the intrinsic dispersion in the spin parameters of the host dark matter halo. The biased collapse plus mergers scenario implies that the specific angular momentum in the stellar components of ETG progenitors at $z\sim 2$ is already close to the local values, in pleasing agreement with observations. All in all, we argue such a behavior to be imprinted by nature and not nurtured substantially by the environment.
\end{abstract}


\section{Introduction}\label{sec_intro}

The relevance of the angular momentum issue in galaxy formation and evolution has been recently reassessed by \cite{Romanowsky2012} and \cite{Fall2013}, who have critically reviewed previous results and have pointed out the still open problems and the main perspectives toward solving them.

In fact, the origin of angular momentum in galaxies has been hotly debated since long times, well before the establishment of the modern cold dark matter (DM) paradigm for structure formation. \cite{Hoyle1949} first pointed out that the tidal field generated by an irregular matter distribution around a proto-galaxy may transfer to it large amount of angular momentum. Such irregular distribution of matter is indeed expected to develop and operate as a consequence of gravitational instability (see \citealt{Sciama1955, Peebles1969, Doroshkevich1970, White1984}). This idea was then successfully applied to compute the angular momentum acquired by galactic DM halos, in the context of the standard cosmological framework (e.g., \citealt{Catelan1996}).

On the observational side, \cite{Takase1967} and \cite{Freeman1970} investigated, for local spiral galaxies, the relationship between the total angular momentum $J_{\star}$ of the stellar disc and the stellar mass $M_\star$, finding a power-law behavior with slope $\approx 7/4$. \cite{Fall1983} pointed out that a more relevant quantity is constituted by the specific angular momentum $j_{\star}=J_{\star}/M_{\star}$, given by the product between a lengthscale and a rotational velocity; he also showed that both spiral and elliptical galaxies follow a relationship $j_{\star}$ vs. $M_{\star}$ with similar slope $\approx 0.6$, but with the former exhibiting systematically larger values of $j_{\star}$ by a factor of $\approx 5$.

The angular momentum of spiral and elliptical galaxies, considered in connection to their structural properties and to the angular momentum of their host DM halos, became soon and still remains a key aspect of galaxy formation and evolution (e.g. \citealt{Efstathiou1979, Efstathiou1980, Davies1983, Mo1998, VanDenBosch2001, Dutton2012, Burkert2016};  for a textbook see \citealt{Mo2010}). \cite{Fall1980} discussed the origin of the rotational properties in disc galaxies within DM halos, by comparing the expectations from the theoretical framework outlined by \cite{White1978} to the available data.

The favored scenario for late-type galaxies (LTGs) envisages that the specific angular momentum of the material forming the disc mirrors that of  the host DM halo (see \citealt{Fall1980, Fall1983, Mo1998}); such an assumption is indeed endorsed by the results of more recent numerical simulations (e.g., \citealt{Governato2007, Zavala2016, Lagos2017}). However, galaxy outflows and tidal stripping have been also advocated in order to rearrange the observed angular momentum in LTGs with different bulge over total mass ratio $B/T$ (see \citealt{Maller2002, Sharma2012, Brook2012, Dutton2012}).

On the other hand, the origin of the low angular momentum measured in early-type galaxies (ETGs) is still open to debate, with a particular focus on the role of merging processes (e.g., \citealt{Hopkins2009}) vs. disc instabilities (e.g., \citealt{Shlosman1993, Noguchi1999, Immeli2004a, Immeli2004b, Bournaud2007}; for a review see \citealt{Bournaud2016}) as possible mechanisms to transfer and/or lose angular momentum.

An original approach to the issue of angular momentum in galaxy formation has been sketched by \cite{Eke2000} and \cite{Fall2002}, starting from the
well known fact that only a fraction $f_{\rm inf}$ of the baryons associated to the DM halo are eventually found in the luminous components of galaxies, namely stars, interstellar medium (ISM), and dust (see \citealt{Persic1992, Fukugita1998}). Then it is reasonable to envisage that only the gas in the inner regions undergoes collapse and fuels star formation, while the outer portions of the galaxy are in a way refrained from forming stars. Since the specific angular momentum of the host DM halo decreases toward the inner regions (e.g., \citealt{Bullock2001}), then the stars mainly formed there should exhibit a lower $j_\star$. \cite{Romanowsky2012} putted forward this `biased collapse' scenario and analyzed its merits and drawbacks.

In the present paper we shall show that actually the infall fraction $f_{\rm inf}$, that provides a quantitative description of the `biased collapse' scenario, can be inferred from observations on the star-formation efficiency and on the chemical abundance of galaxies. The data indicate that the fraction $f_{\rm inf}$ is appreciably different for ETGs and LTGs, implying that the two galaxy types occupy distinct loci in the specific angular momentum vs. stellar mass diagram; as a consequence,
ETGs and LTGs are found to have retained in their stellar component
a different fraction $f_j$ of the angular momentum initially associated
to the infalling baryons. We shall estimate such
quantities and discuss how to physically interpret them in the light of a
a biased collapse plus merger scenario.

The plan of the paper is the following. After a brief presentation of the argument (Sect.~\ref{sec_initial}), in Sect.~\ref{sec_finf} we show how to infer a robust estimate of the infalling gas fraction as function of the stellar mass for both ETGs and LTGs, by exploiting their observed star-formation efficiency and metal abundance. Sect.~\ref{sec_fstar_z} is devoted to present and summarize the available data on star-formation efficiency and metallicity in ETGs and LTGs. The infalling gas fraction and its impact on the specific angular momentum of both galaxy types are investigated in Sect.~\ref{sec_finf_jstar}. In Sect.~\ref{sec_discussion} we discuss our results and compare them with recent observational data and numerical simulations. Sect.~\ref{sec_sum} summarizes our key findings.

Throughout this work we adopt the standard flat cosmology from \cite{Planck2016} with round parameter values: matter density $\Omega_M = 0.31$, baryon density $\Omega_b = 0.05$, Hubble constant $H_0 = 100\, h$ km s$^{-1}$ Mpc$^{-1}$ with $h = 0.67$, and mass variance $\sigma_8 = 0.83$ on a scale of $8\, h^{-1}$ Mpc. Stellar masses and star formation rates (or luminosities) of galaxies are evaluated assuming the \cite{Chabrier2003} initial mass function (IMF).

\section{The initial specific angular momentum of inflowing gas}\label{sec_initial}

The galaxy angular momentum acquired by protogalaxies is classically presented in terms of the dimensionless spin parameter
\begin{equation}\label{eq_lambda}
\mathrm{\lambda} \equiv \frac{J\,|E|^{1/2}}{G\, M^{5/2}} ,
\end{equation}
which is a combination of basic galactic physical quantities, namely the total angular momentum $J$, the total energy $E$, and the total mass $M$ (DM and baryons; see \citealt{Peebles1969, Peebles1971}). The distribution of the spin parameter as function of mass, redshift, and environment has been studied both with analytic approximations as well as numerical simulations (e.g., \citealt{Barnes1987, Maccio2007, Bett2007, Rodriguez-Puebla2016}). The emerging picture envisages that the halo spin parameter exhibits a lognormal distribution with average $\langle \lambda \rangle\approx 0.035$ and dispersion $\sigma_{\log\lambda} \approx 0.25$ \citep{Rodriguez-Puebla2016}, nearly independent of mass and redshift, but somewhat dependent on environment (e.g., \citealt{Bett2007, Maccio2007, Maccio2008, Shi2015}). After \cite{Romanowsky2012}, we can define the specific angular momentum $j\equiv J/M$ of a spherically symmetric DM halo with mass distribution following a NFW profile (\citealt{Navarro1996, Navarro1997}) extended out to the conventional virial radius $r_{\rm vir}$:
\begin{equation}\label{eq_jvir}
j(r_{\rm vir})\approx 4.2\times 10^4 \lambda \left(\frac{M_{\rm vir}}{10^{12}\,M_{\sun}}\right)^{2/3}\, E(z)^{-1/6}\,\, {\rm km\, s^{-1}\, kpc},
\end{equation}
where $E(z)\equiv \Omega_{\Lambda}+\Omega_{M}\, (1+z)^3$. Note that the redshift dependence is weak; for instance, a halo at $z\approx 2$ features a momentum $j(r_{\rm vir})$ lower by a relatively small factor $\approx 1.4$ than a halo with the same mass at $z=0$.

\cite{Barnes1987} and \cite{Bullock2001} pointed out via $N-$body simulations that the radial distribution of the halo specific angular momentum is well described by a power-law with exponent $s\approx 1$, i.e.,
\begin{equation}
\label{eq_jr}
j(r)=j(r_{\rm vir})\, \left[\frac{M(\le r)}{M(\le r_{\rm vir})}\right]^s
\end{equation}
implying that the inner regions of halos exhibit lower specific angular momentum than outer ones. In Appendix A we exploit state-of-the-art, high-resolution $N-$body simulations to derive the distribution of the parameter $s$ as a function of mass and redshift (see also Figs.~\ref{fig_JM_Prof} and \ref{fig_sMhz}).

Next we assume that initially the baryonic mass follows the same radial distribution of the DM with ratio $f_b\equiv M_b/M_{\rm vir}=\Omega_b/\Omega_M$; thus also the distribution of specific angular momenta for the baryonic gas $j_b(r)$ and for the DM $j(r)$ mirrors each other, i.e., $j_b(r)=j(r)$. However, it could happen that only a fraction $f_{\rm inf}$ of the baryons associated with the galaxy halo is able to cool down and flow inward to reach the inner regions where most of the star formation occurs. Then such baryons are expected to feature a specific angular momentum lower than $j_{\rm vir}$. More in detail, after Eq.~(\ref{eq_jr}) the fraction of baryons involved in the formation of the galaxy $f_{\rm inf}\equiv M_{\rm inf}/f_b\, M_{\rm vir}=M_b(\le r_{\rm inf})/M_b(\le r_{\rm vir})\le 1$ has an initial specific angular momentum
\begin{equation}\label{eq_jinf1}
j_{\rm inf}=j(r_{\rm vir})\, f_{\rm inf}^s~.
\end{equation}
Note that this equation is very similar in spirit to Eq. (14) by \cite{Fall1983}, who advocated tidal stripping as a possible mechanism to prevent baryon collapse from the outer regions of halos hosting ETGs.

As we shall see below (see Sect. \ref{sec_fstar_et} and \ref{sec_fstar_lt}), the halo mass for galaxies endowed with stellar mass $M_{\star}$ can be estimated via various techniques. The outcome is standardly expressed in terms of the star-formation efficiency $f_{\star} \equiv M_{\star}/f_b\, M_{\rm vir}$ as a function of the stellar mass $M_\star$. Plugging the definition of $f_\star$ in Eqs.~(\ref{eq_jvir}) and (\ref{eq_jinf1}), we can write down the intrinsic angular momentum of the inflowing gas as ùfunction of the stellar mass and star-formation efficiency
\begin{equation}\label{eq_jinf}
j_{\rm inf}\approx 3.1\times 10^4\, \lambda\,  f_{\star}^{-2/3}\, \left (\frac{M_{\star}}{10^{11}\,M_{\sun}}\right)^{2/3}\,f_{\rm inf}^s\,  E(z)^{-1/6}~ {\rm km\, s^{-1}\, kpc}~.
\end{equation}
The above formula differs from Eq.~(15) of \cite{Romanowsky2012} in two respects: (i) we introduce the dependence on redshift (see also \citealt{Burkert2016}); (ii) we focus on the specific angular momentum of the infalling gas. By comparing the observed $j_{\star}$ to $j_{\rm inf}$, we aim to determine the fraction $f_j\equiv j_\star/j_{\rm inf}$ of the initial specific angular momentum retained by the stellar component (see \citealt{Romanowsky2012}).

The next section is devoted to develop a method aimed at estimating the infalling baryon fraction $f_{\rm inf}$ from the observed star-formation efficiency and metal abundance for both ETGs and LTGs.

\section{Fraction of inflowing gas from stellar efficiency and metal abundance}\label{sec_finf}

In the local Universe, most of the baryonic mass within the central region (size $\la 10-20$ kpc) of galaxies comprises three main components: stars, dust, and interstellar medium (ISM). An additional, diffuse component of warm/hot gas, often dubbed circumgalactic medium (CGM; e.g., \citealt{Tumlinson2011}), pervades a much larger volume up to hundreds kpc. These components descend from the diffuse  gas of mass $M_b$ associated with the galactic halo at the epoch of halo virialization. A portion of (or all) the gas cools down from the initial virial temperature, allowing star formation to occur especially in clumpy regions, and chemical enrichment of the galactic components to proceed. A fraction of the cooled gas can be eventually expelled from the central regions by energy/momentum feedback associated to supernovae explosions/stellar winds and outbursts from the central active galactic nucleus (AGN). These feedbacks, depending on the history of star formation and AGN accretion, can be so efficient as to quench star formation and to forbid further cooling of the hot/warm gas (see \citealt{White1991, Bressan1998, Cole2000}); this is particularly true for AGNs, that are indeed expected to originate large-scale outflows (see \citealt{Granato2004, DiMatteo2005, Lapi2006}). A statistical evidence of the latter can be recognized in the chemical enrichment of the intracluster medium (e.g., \citealt{Leccardi2010, Bohringer2014}).

The total mass of the observed baryonic components, namely the mass in stars $M_{\star}$ (including stellar remnants), ISM $M_{\rm ISM}$, dust $M_{\rm dust}$, and CGM $M_{\rm CGM}$, should not exceed the mass $M_b$ of the baryons associated with the galaxy halo. On the other hand, the balance of the baryonic mass that cool and infall toward the central regions $M_{\rm inf}$, the mass of the baryons still in the galaxy $M_{\rm gal}$=$M_{\star}+M_{\rm ISM}+M_{\rm dust}$, and the mass $M_{\rm out}$ of the gas expelled from the central regions by feedback mechanisms can be written as
\begin{equation}\label{eq_minf}
M_{\rm gal} \equiv M_{\rm inf}-M_{\rm out}~;
\end{equation}
note that the CGM does not enter the galaxy mass balance. As for the budget of metals, \cite{Fukugita2004} have shown that most of them are locked up into compact objects like white dwarfs, neutron stars, and stellar mass black holes. However, here we are interested in the budget of the metals produced by stars but not locked up into their compact remnants; we denote them as \emph{accessible} metals. Observations of stellar metallicity in galaxies refer essentially that of main sequence stars, after proper luminosity weighting.

In order to evaluate the accessible metals produced in a galaxy,
a relevant quantity is constituted by the true metal yield $y_Z$ of a single stellar population; here we adopt the classic definition of $y_Z$ that includes a normalization to $1-R$, where $R$ is the return fraction of gaseous material from the formed stars (e.g., \citealt{Vincenzo2016}).
In the following we assume instantaneous recycling, but we have checked with detailed chemical evolution models that this is indeed a good approximation in our context (see also \citealt{Feldmann2015, Vincenzo2016}). Note that $y_Z$ depends on the assumed Chabrier IMF, and mildly on the chemical composition of the stars; however, for our purposes this is a second order effect, so that we just exploit average yields appropriate for reasonable chemical abundances (e.g. \citealt{Peeples2014, Feldmann2015, Vincenzo2016}). Under these assumptions the total mass of accessible metals produced by stars is then
\begin{equation}
\label{eq_mz}
M_Z=y_Z\, M_{\star}~,
\end{equation}
where $y_Z\approx 0.069$ applies for a Chabrier IMF \citep{Krumholz2012}.

The budget of accessible metals inside the galaxy reads
\begin{equation} \label{eq_mzgal}
M_{Z, \rm gal}=M_{Z,\star}+M_{Z,\rm ISM}+M_{Z,\rm dust}=\langle Z_{\star}\rangle\,  M_{\star} +\langle Z_{\rm ISM}\rangle\, M_{\rm ISM}+M_{\rm dust}~,
\end{equation}
where we set $M_{Z,\star}=\langle Z_{\star}\rangle\, M_{\star}$, $M_{Z,\rm ISM}=\langle Z_{\rm ISM}\rangle\, M_{\rm ISM}$ and $M_{Z,\rm dust}=M_{\rm dust}$, i.e. $\langle Z_{\rm dust}\rangle=1$. The metal mass conservation implies (see e.g. \citealt{Peeples2014})
\begin{equation} \label{eq_mz1}
M_Z=M_{Z, \rm gal}+M_{Z,\rm out}~,
\end{equation}
where $M_{Z,\rm out}=\langle Z_{\rm out}\rangle\,  M_{\rm out}$ is the mass of metals expelled from the galaxy and disseminated in the CGM and IGM (see \citealt{Peeples2014}). The above equation assumes that: (i) the cool gas inflowing from the galactic halo has a negligible metal content; (ii) outflowing mass and metals do not fall back at later times (i.e., no circulation due to a galactic fountain). We shall discuss in Appendix B that relaxing such assumptions do not alter appreciably our results and conclusions.

Replacing $M_{Z, \rm gal}$ and $M_{\rm out}$ after Eqs.~(\ref{eq_mzgal}) and (\ref{eq_minf}), we get
\begin{equation} \label{eq_mz2}
y_Z\, M_{\star}= \langle Z_{\star}\rangle \, M_{\star} + \langle Z_{\rm ISM}\rangle\, M_{\rm ISM}+M_{\rm dust} + \langle Z_{\rm out}\rangle \, (M_{\rm inf}-M_{\star}-M_{\rm ISM}-M_{\rm dust})~.
\end{equation}
Then we express the average metal abundance of the outflowing gas in terms of the stellar metallicity via the parameter $\zeta \equiv \langle Z_{\rm out}\rangle/\langle Z_{\star}\rangle$ and insert the star-formation efficiency $f_\star\equiv M_\star/f_b\,M_{\rm vir}$ and infall fraction $f_{\rm inf}\equiv M_{\rm inf}/f_b\,M_{\rm vir}$, to obtain
\begin{equation}\label{eq_finf}
f_{\rm inf}= f_{\star}\, \left( \frac{y_Z}{\zeta\, \langle Z_{\star}\rangle}  -\frac{M_{Z, \rm gal}}{\zeta\,  \langle Z_{\star}\rangle\, M_{\star}} +
\frac{M_{\rm gal}}{M_{\star}}\right)~.
\end{equation}

We can infer $\zeta$ from quite general arguments. In the case of feedback originated by stellar winds and supernova explosions the outflow rate is proportional to the star formation rate $\dot{M_{\rm out}}\approx\epsilon _{\rm out}\, \dot{M_{\star}}$, where $\epsilon_{\rm out}$ is the mass loading factor; then $M_{\rm out}(\tau)\approx \epsilon_{\rm out}\, M_{\star}(\tau)$ holds at any galactic age $\tau$ (e.g. \citealt{Feldmann2015}), implying that both the stellar metallicity $Z_{\star}$ and the outflows metallicity $Z_{\rm out}$ can be computed as
\begin{equation}\label{eq_zstarout}
Z_{X}(\tau) = {1\over M_{X}(\tau)}\, \int_0^{\tau}{\rm d}\tau'~Z_{\rm gas}(\tau')\, \dot M_{X}(\tau')~,
\end{equation}
with $X=\star$ or $X={\rm out}$. As a consequence the metallicity of the stars and of the outflows are quite close to each other $Z_{\rm out}(\tau)\approx Z_{\star}(\tau)$. Therefore for galaxies with outflows dominated by stellar feedback, e.g. LTGs, $\zeta\approx 1$ applies.

On the other hand, the effect of the AGN feedback, relevant in the case of ETGs, can simply be described as an abrupt quenching of the star formation, where most of the gas is assumed to be removed.  If the feedback occurs at time $\tau_{\rm AGN}$, then the metallicity reads
\begin{equation}\label{eq_zout}
Z_{\rm out}= \frac {M_{Z,\rm out}}{M_{\rm out}}=\frac{Z_{\star}(\tau_{\rm AGN})\, \epsilon_{\rm out} M_{\star}(\tau_{\rm AGN}) + Z_{\rm gas}(\tau_{\rm AGN})\,  M_{\rm gas}(\tau_{\rm AGN})}{\epsilon_{\rm out}\, M_{\star}(\tau_{\rm AGN}) + M_{\rm gas}(\tau_{\rm AGN})}~.
\end{equation}
Since the gas metallicity $Z_{\rm gas}$ is increasing with time, Eq.~(\ref{eq_zstarout}) implies that the metal abundance of the stars is lower than that of the gas for small galactic age $\tau\ll 10^8$ yr, but they converge $Z_{\rm gas}(\tau)\approx 1.1\, Z_{\star}(\tau)$ after a few $10^8$ yr. As a result $Z_{\rm out}\approx Z_{\star}$ holds also in the case of AGN feedback. Summing up, we conclude that $\zeta\approx 1$ applies for both ETGs and LTGs.

\section{Star-formation efficiency and metallicity of ETGs and LTGs}\label{sec_fstar_z}

In this Section we examine the star-formation efficiency (or equivalently the stellar to halo mass ratio) and the metal abundance in ETGs and LTGs.

The host halo mass of galaxies has been investigated exploiting different observational approaches and theoretical assumptions; the more common techniques involve satellites kinematics, weak gravitational lensing, and abundance matching. Satellite kinematics and weak lensing offer the important opportunity of studying separately ETGs and LTGs (e.g. \citealt{More2011, Wojtak2013, Hudson2015, Velander2014, Mandelbaum2016}). In particular, the weak lensing has been exploited to investigate large samples of galaxies via stacking techniques, even at significant redshift $z\la 0.7$ (e.g., \citealt{Hudson2015}). Abundance matching also provides insights on the galaxy-to-halo mass ratio at substantial redshift (e.g., \citealt{Shankar2006, Moster2013, Behroozi2010, Behroozi2013, Aversa2015, Huang2017}), although the separation between galaxy types is more challenging.

\begin{figure}
\centering
\includegraphics[width=0.55\linewidth]{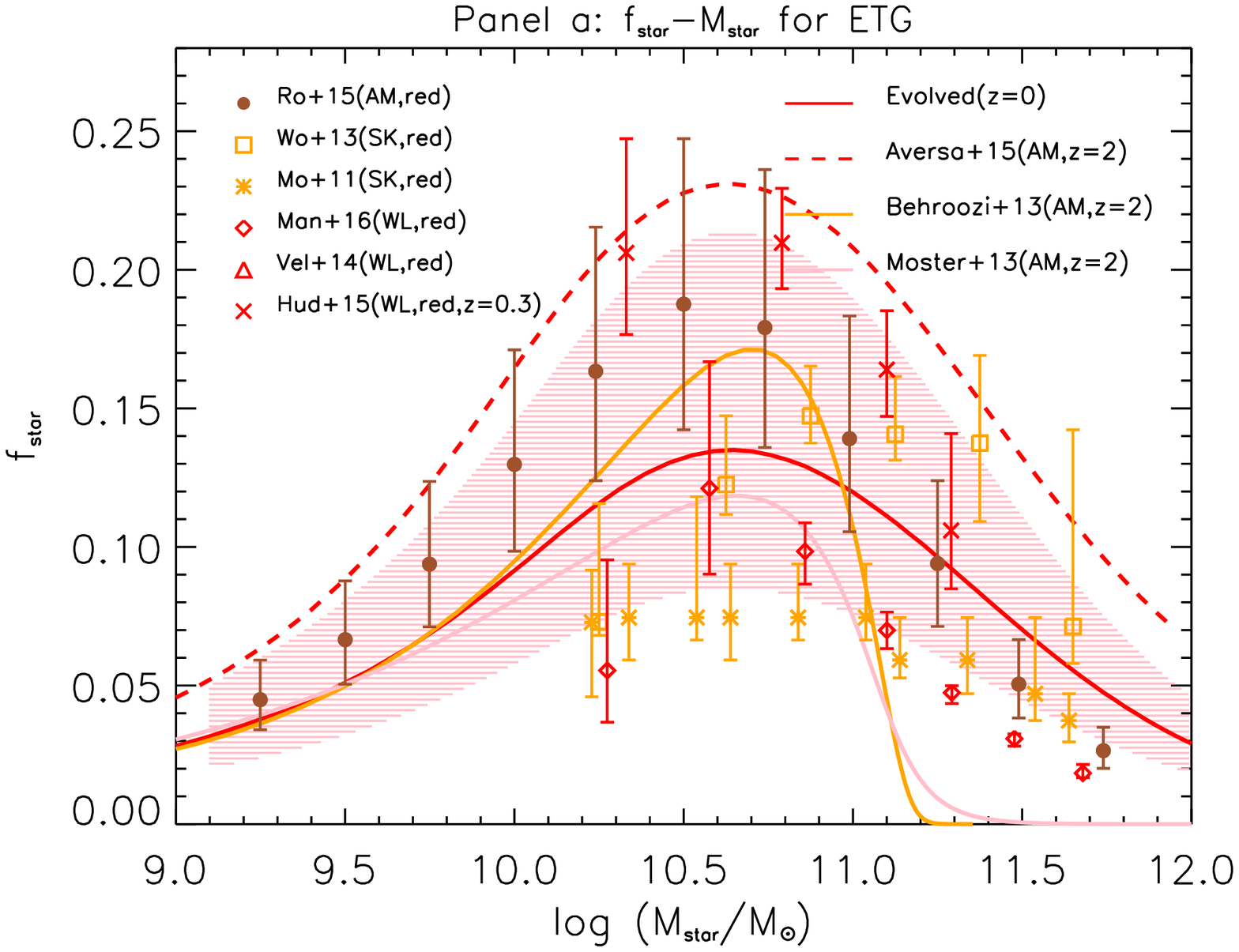}
\includegraphics[width=0.55\linewidth]{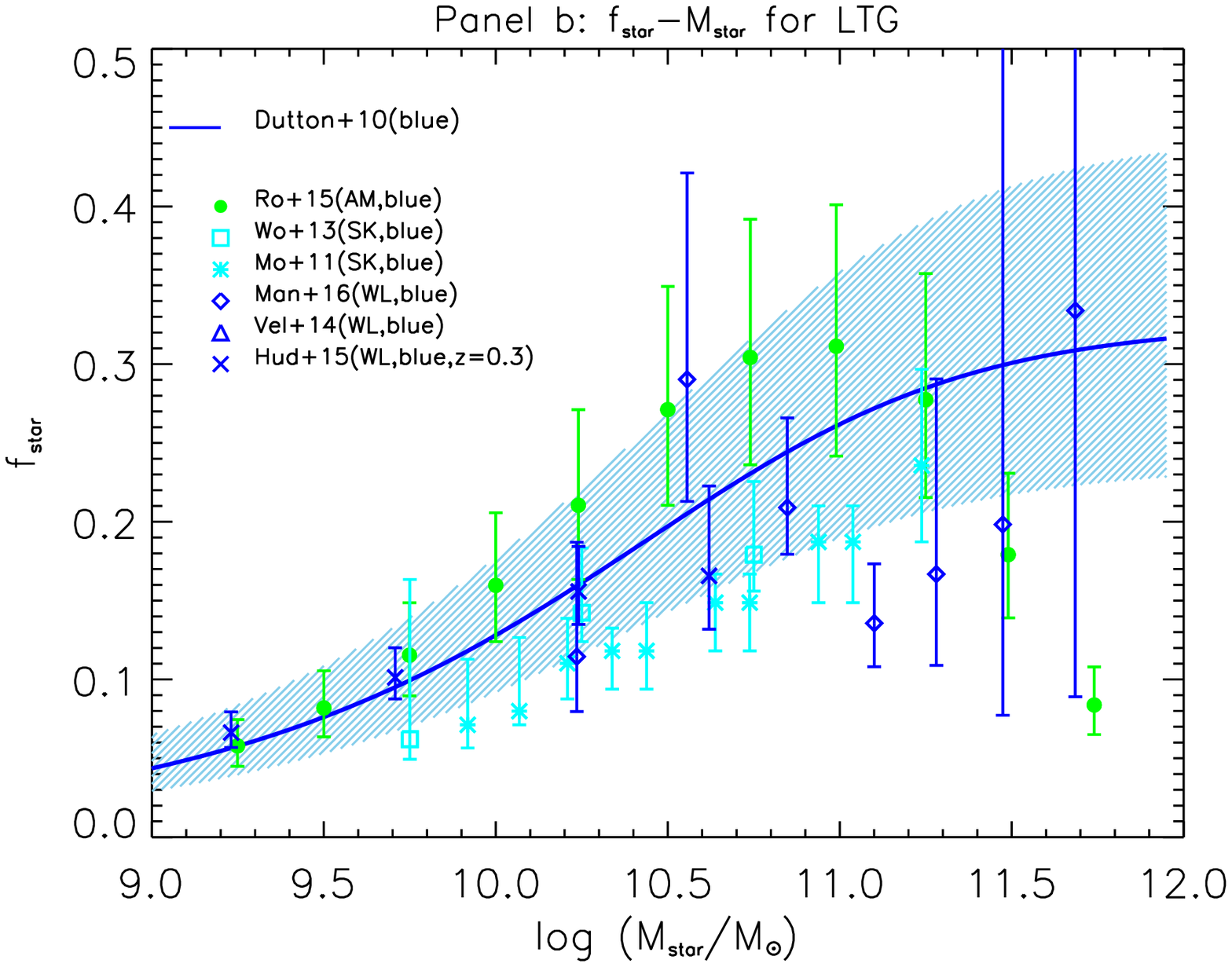}
\caption{Star formation efficiency $f_{\star}$ versus stellar mass $M_{\star}$ for ETGs (top panel) and LTGs (bottom panel).  Top panel: red dashed line represents the relationship at $z=2$ for ETGs inferred from \cite{Aversa2015} via the abundance matching technique, while the red solid line is the same relationship evolved to $z=0$ (see details in Sect.~\ref{sec_fstar_et}), with the red shaded area showing the 1$\sigma$ uncertainty. The orange and pink lines are the abundance matching results at $z=2$ from \cite{Behroozi2013} and \cite{Moster2013}, respectively. Filled circles are the abundance matching data for red galaxies at $z=0$ from \cite{Rodriguez-Puebla2015}. Other datapoints are weak lensing or satellite kinematic measurements in the local Universe from \cite{Wojtak2013}, \cite{More2011}, \cite{Mandelbaum2016}, \cite{Velander2014} and \cite{Hudson2015} at $z=0.3$. Bottom panel: blue solid line represents the $f_{\star}$ vs. $M_{\star}$ relation for LTGs from \cite{Dutton2010}, with the blue shaded area indicating the 1$\sigma$ uncertainty. Data are from the weak lensing and satellite kinematic observations cited above, but for blue galaxies.}
\label{fig1_fstar}
\end{figure}

\subsection{Star-formation efficiency of ETGs }\label{sec_fstar_et}

In Fig.~\ref{fig1_fstar} (top panel) we present the star-formation efficiency of ETGs as function of their stellar mass for relatively local samples at $z\la 0.3$. Data are from recent estimates based on satellite kinematics (\citealt{More2011, Wojtak2013}), weak lensing
(\citealt{Hudson2015, Velander2014, Mandelbaum2016}), and abundance matching (\citealt{Rodriguez-Puebla2015}). Most data refer to the central/brightest red galaxy of a halo, possibly corrected for the contribution from satellites. This procedure is quite complex and can significantly contribute to the observed scatter of about $0.2$ dex (see \citealt{Behroozi2013, Reddick2013, Huang2017}), as shown by the red shaded area in the top panel of Fig.~\ref{fig1_fstar}.

Now we turn to the problem of estimating the star-formation efficiency at the reference redshift/epoch when most ($\ga 70\%$) of the stars have been formed in the ETG progenitors. The notions that ETGs are quite old systems (formation redshift $z\ga 1$) and that they formed in a relatively short timescale $\la 1$ Gyr is time honored (e.g., \citealt{Bower1992, Thomas2005, Thomas2010}; for a review see \citealt{Renzini2006}). This is strongly supported by recent archeological studies on massive passively evolving galaxies at substantial redshift $z\la 1$ , which shows that they formed most of their stars at $z\sim 1.5-2$ (e.g. \citealt{Trujillo2011, Onodera2015, Lonoce2015, Citro2016, Siudek2017, Kriek2016, Gallazzi2006, Gallazzi2014, Choi2014, Glazebrook2017}). Even lower mass ETGs formed mostly at $z\sim 1$, as pointed out by \cite{Siudek2017}. We further notice that the cosmic stellar mass density increased by $\approx 40\%$ from $z\approx 1$ to the present (see \citealt{Madau2014, Aversa2015}); this increase corresponds to the present-day fraction of stellar mass density contributed by disc dominated galaxies, including $Sa$ (e.g., \citealt{Moffett2015}).

Investigations on the fraction of close galaxy pairs and of galaxies with disturbed morphologies in large catalogs (e.g., \citealt{Man2016}) indicate that the mass growth of massive galaxies $M_{\star}\ga 7\times 10^{10}\, M_{\sun}$ is constrained within a factor $\approx 1.5-2$ in the redshift interval $z\sim 0.1-2.5$. Limited mass evolution $\Delta \log M_{\star}\approx 0.16\pm 0.04$ is also confirmed for a sample of quiescent galaxies at redshift $z\la 1.6$ by \cite{Belli2014}.

In the following we assume for ETGs a reference formation (when most $\ga 70\%$ of the stars have been formed) redshift $z\approx 2$ and an average stellar mass increase of $50\%$ since then. Since the stellar mass function at $z\approx 2$ is mainly dominated by the ETG progenitors, it's reasonable for us to exploit the abundance matching technique applied to galaxies at $z\approx 2$ in order to derive an estimation of the star-formation efficiency in ETG progenitors.

In Fig.~\ref{fig1_fstar} (top panel) we present the outcome of the abundance matching at $z\approx 2$ between the stellar and halo mass functions computed by \cite{Aversa2015}. The results from \cite{Moster2013} and \cite{ Behroozi2013} are also shown for comparison. The resulting star-formation efficiencies differ by no more than a factor of $2$. A relavent check on the efficiency can be done by comparing the estimate at $z\approx 2$ to the low redshift estimates based on weak lensing and on satellites kinematics. Evolution in both halo and stellar mass must be taken into account. As for the stellar mass change, we assume an increase of $50\%$ as mentioned above.

As for the halo mass evolution, it has been computed via $N-$body simulations by \cite{McBride2009} and \cite{Fakhouri2010}, and via the excursion set approach by \cite{Lapi2013}, with concordant results. The main progenitor of a present-day halo with virial mass $M_{\rm vir}$ evolves from $z\approx 2$ to the present as
\begin{equation}\label{eq_mh}
M_{\rm vir}(z=0)\approx 4.0\, M_{\rm vir}(z=2)\, \left[\frac{M_{\rm vir}(z=0)}{10^{14}\, M_{\sun}}\right]^{0.12}~.
\end{equation}
The solid red line in Fig.~\ref{fig1_fstar} shows how the star-formation efficiency estimated by \cite{Aversa2015} at $z\approx 2$ evolves toward $z\approx 0$ along the assumed evolutionary pattern for DM and stellar mass. The agreement with local data derived from weak gravitational lensing and satellites kinematics is nice. We also checked that the proposed evolution is similar to that inferred by \cite{Hudson2015} for red galaxies with stellar mass $M\ga 2\times 10^{10}\, M_{\sun}$ (in our framework ETGs) between $z\approx 0.7$ and $z\approx 0.3$. Therefore, we adopt the estimation from \cite{Aversa2015} as star formation efficiency for $z=2$ ETG progenitors.

\subsection{Metal abundance of ETGs}\label{sec_z_et}

In order to derive the inflowing gas fraction $f_{\rm inf}$ from Eq.~(\ref{eq_finf}), not only the star-formation efficiency $f_\star$ but also the stellar metallicity $Z_{\star}$ at $z\approx 2$ is needed.

As for the stellar metallicity of ETGs, we adopt the relationship $Z_{\star}$ vs. $M_{\star}$ proposed by \cite{Gallazzi2006} for a local $z\la 0.2$ galaxy sample, with its $1\sigma$ scatter of $0.12$ dex (red line and red shaded area in the bottom panel of Fig.~\ref{fig2_zfstar}). There is evidence that after the main burst of star formation, the metal abundance of stars in ETGs keeps practically constant (e.g., \citealt{Citro2016, Gallazzi2014, Choi2014, Siudek2017}), as confirmed from high redshift observations of passively evolving galaxies (see \citealt{Lonoce2015, Kriek2016}). Therefore we reasonably assume that the present-day metallicity of massive ETGs was already in place at redshift $z\approx 2$. For ETGs we also neglect both dust and ISM in the mass and metals budget.

\subsection{Star formation efficiency of LTGs}\label{sec_fstar_lt}

By comparing the panels of Fig.~\ref{fig1_fstar}, it is apparent that local LTGs exhibit a larger star-formation efficiency than ETGs. In particular, at high stellar masses LTGs appear more efficient by a factor of $1.5-2$ (see \citealt{Dutton2010, More2011, Wojtak2013, Velander2014, Rodriguez-Puebla2015, Mandelbaum2016}). Despite the large scatter of the data, a higher efficiency for LTGs is found from several samples, independent of whether the halo mass is derived via abundance matching or via weak lensing. In the bottom panel of Fig.~\ref{fig1_fstar} we illustrate the fit to the data by \cite{Dutton2010}, with its associated 1$\sigma$ uncertainty shown by the blue shaded area in the bottom panel of Fig.~\ref{fig1_fstar}.

At variance with ETGs, \cite{Hudson2015} show that the relationship between efficiency and stellar mass does not appreciably evolve between $z\approx 0.7$ and $z\approx 0.3$. A straightforward interpretation is that in LTGs the star formation and DM accretion go in parallel along cosmic times. In the following we assume that the star-formation efficiency vs. stellar mass relationship in LTGs stays almost constant, close to the present-day value, along the period of disc formation.

\subsection{Metal abundance of LTGs}\label{sec_z_lt}

In the case of LTGs the mass in ISM and in dust are no more negligible and, as a consequence, they can contribute significantly to the global galaxy metal abundance. The amounts of stars, ISM, dust, and their metal abundance have been presented by \cite{Peeples2014}. We adopt their relationships with the associated scatter, and defer the reader to their paper for details. Note that the stellar metallicity measurements still retain an appreciable  uncertainty (\citealt{Gallazzi2005, Goddard2017}), especially for low mass LTGs (cf. bottom panel of Fig.~\ref{fig2_zfstar}). Another caveat concerns  the metal mass in the ISM, which includes only cold gas in the analysis of \cite{Peeples2014}; the mass and metals in warm ionized gas could be as large as those in the cold gas (see \citealt{Sembach2000, Haffner2009, Peeples2014}). Then we checked that doubling the ISM mass and metals affects only marginally our results; e.g., the infall fraction $f_{\rm inf}$ (cf. Sect. \ref{sec_finf_jstar}) changes by no more than $10\%$.

We recall that LTGs are still forming stars in their discs, at exponentially declining rates (e.g., \citealt{Chiappini1997}). This implies
the metallicity to increase along cosmic times; the median increase since $z\approx 0.7$ to the present has been estimated by \cite{Gallazzi2014} to be $\la 0.12$ dex, which is comparable to the uncertainties in the metallicity estimates (see \citealt{Gallazzi2005, Peeples2014}).

\section{Estimated fraction of inflowing gas and specific angular momentum}\label{sec_finf_jstar}

\begin{figure}
\centering
\includegraphics[width=0.8\linewidth]{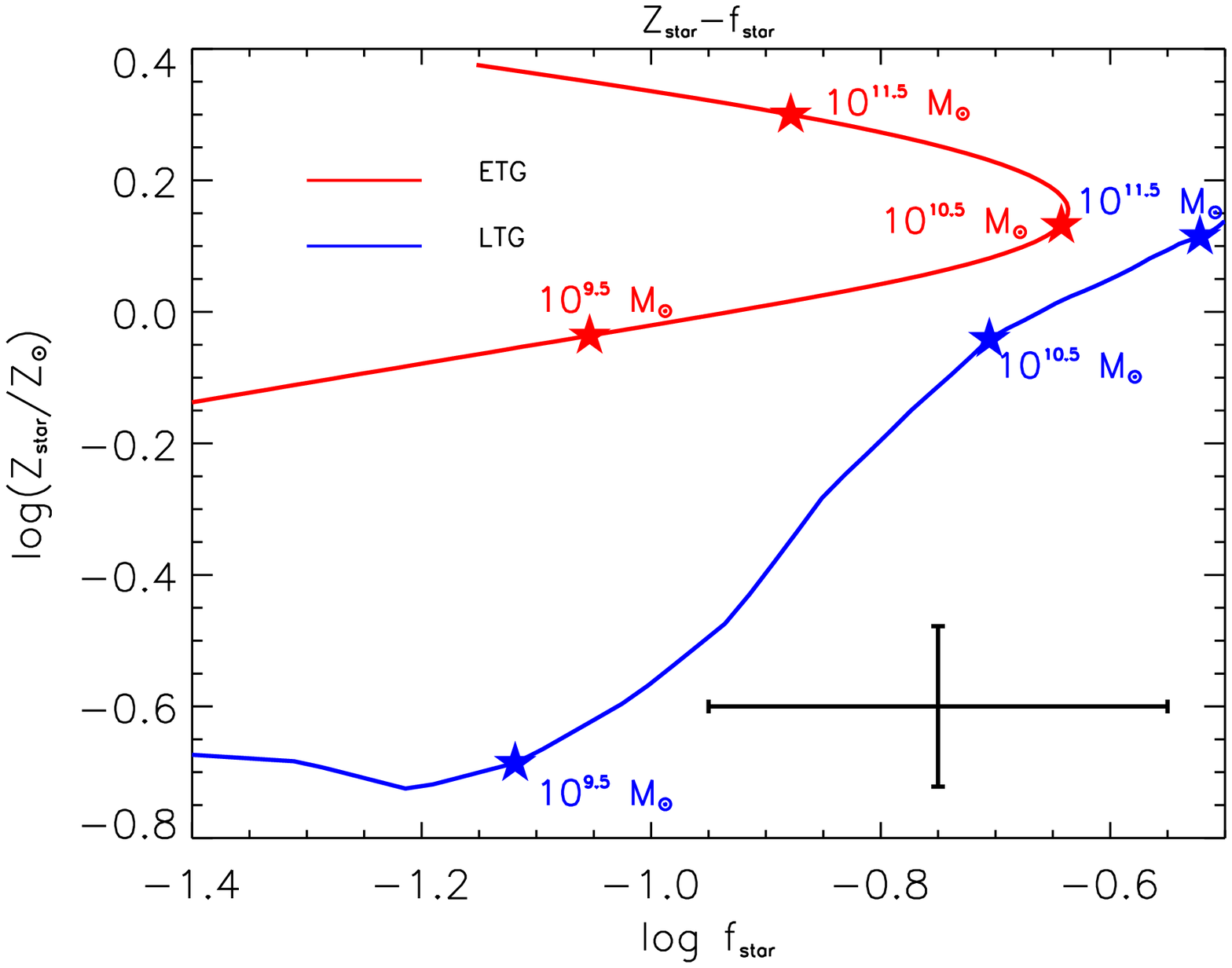}
\includegraphics[width=0.8\linewidth]{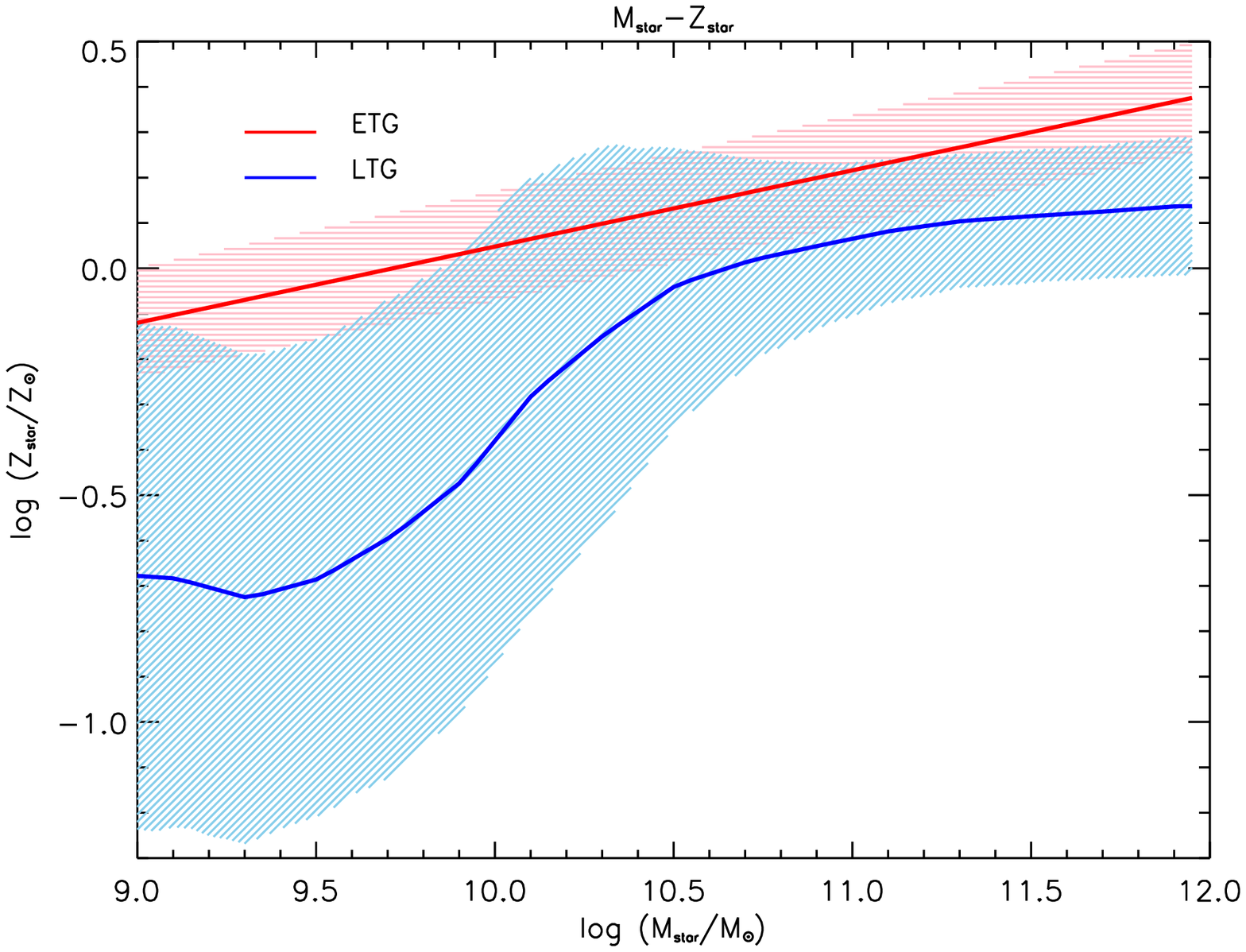}
\caption{ Top panel: The stellar metallicity $Z_{\star}$ (in units of the solar value $Z_\sun=0.015$) plotted against star-formation efficiency $f_{\star}$ for ETGs (red) and LTGs (blue). The stars highlight the position on the curves for galaxies with stellar masses $M_\star\approx 10^{9.5}-10^{10.5}-10^{11.5}\, M_\sun$. The error bars in the right corner indicates the typical uncertainty in the measurements of $Z_{\star}$ and $f_{\star}$. Bottom panel: The stellar metallicity $Z_{\star}$ versus $M_{\star}$ for ETGs (red) and LTGs (blue), taken from \cite{Gallazzi2005} and \cite{Gallazzi2006}. The red and blue shaded areas show the 1$\sigma$ uncertainty.}
\label{fig2_zfstar}
\end{figure}

\begin{figure}
\centering
\includegraphics[width=0.8\linewidth]{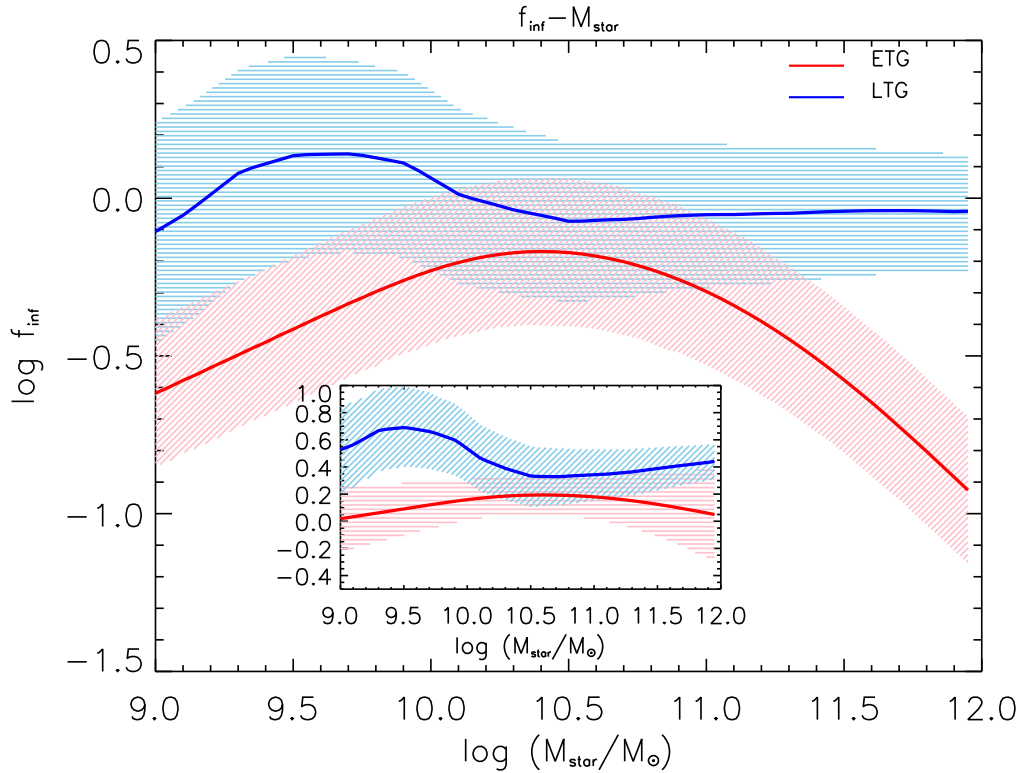}
\caption{The inferred baryon infalling fraction $f_{\rm inf}$ for ETGs (red line) and LTGs (blue). The colored shaded areas indicate the 1$\sigma$ uncertainty calculated by taking into account the scatter of the parameters entering Eq.~(\ref{eq_finf}). In the inset, we plot for ETGs and LTGs the quantity $y_Z\,f_{\star}^{-2/3+s}\, Z_{\star}^{-s}\, M_{\star}^{0.15}$ together with its $1\sigma$ uncertainty (colored shaded areas; see text for details).}
\label{fig3_finf}
\end{figure}

\begin{figure}
\centering
\includegraphics[width=0.6\linewidth]{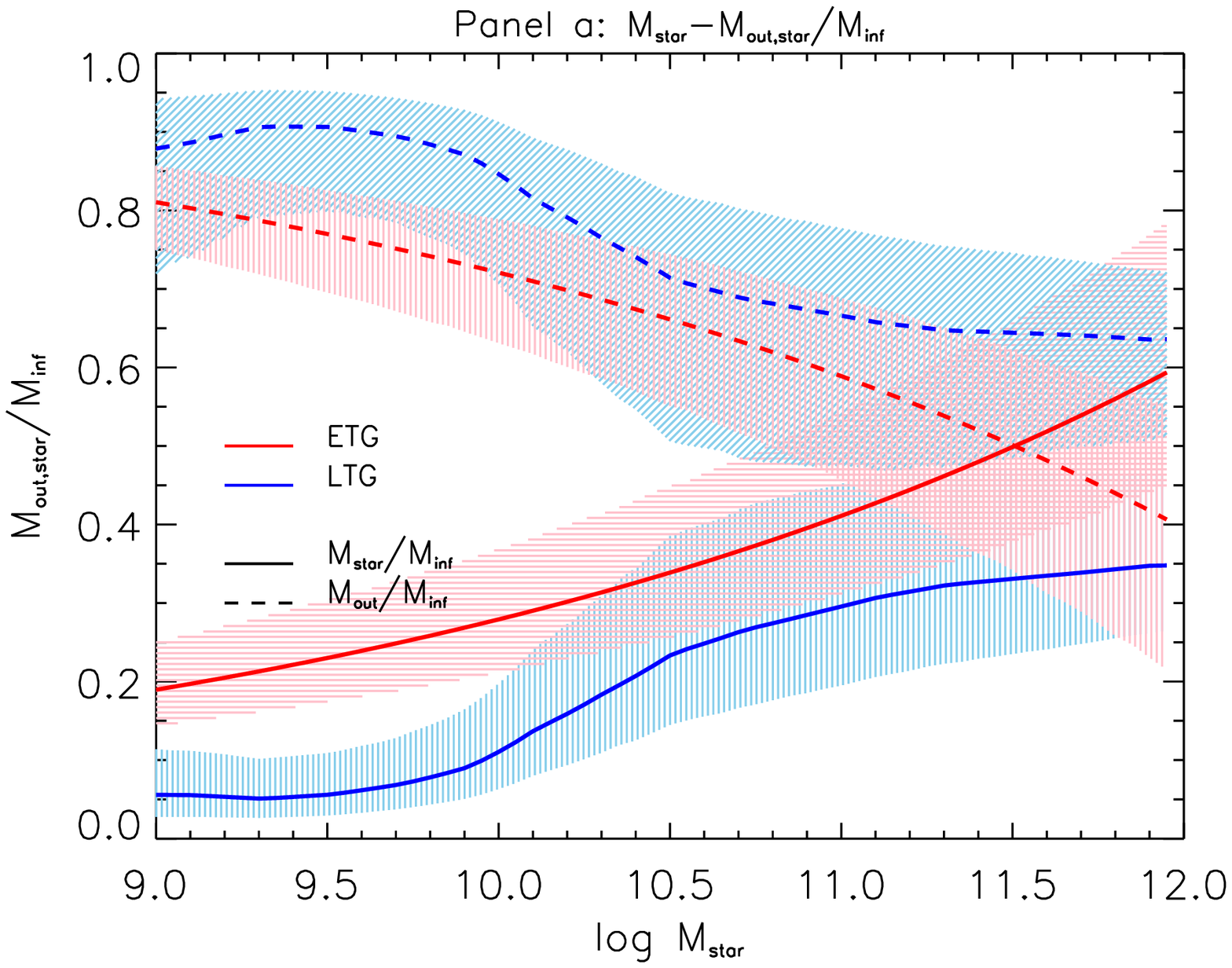}
\includegraphics[width=0.6\linewidth]{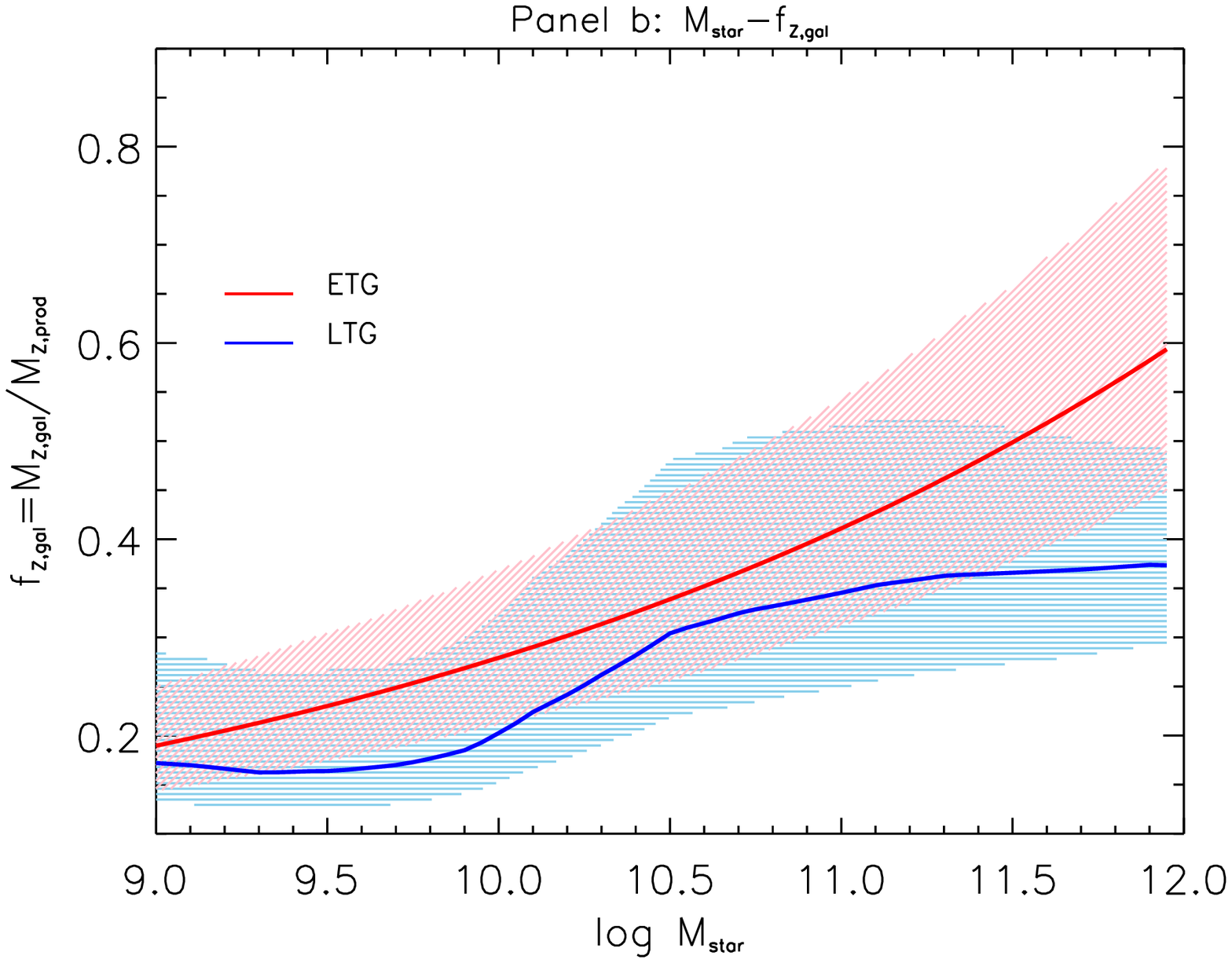}
\caption{Top panel: stellar mass fraction $M_{\star}/M_{\rm inf}$ (solid lines) and ejected out mass fraction $M_{\rm out}/M_{\rm inf}$ (dashed lines) for ETGs (red) and LTGs (blue). Bottom panel: fraction of metals retained in galaxies $f_{Z,\rm gal}\equiv M_{Z,\rm gal}/M_{Z,\rm prod}$. The shaded areas show explicitly the 1$\sigma$ uncertainties of the estimates.}
\label{fig4}
\end{figure}

Fig.~\ref{fig2_zfstar} highlights that ETGs and LTGs occupy different loci in the $f_{\star}$ vs. $Z_{\star}$ plane. At given metallicity the efficiency is  higher for LTGs; on the other hand, ETGs feature higher metallicity even if the efficiency is small. These observational results impact directly on the fraction $f_{\rm inf}$ of gas flowing into the central regions (cf. Eq.~\ref{eq_finf}).

In Fig.~\ref{fig3_finf} the fraction $f_{\rm inf}$ is plotted against the stellar mass; the shaded areas reflect the uncertainties in chemical abundance and stellar efficiency. In the case of LTGs the resulting fraction $f_{\rm inf}\approx 0.9-1.3$ is very close to $1$, except in a limited mass range $M_\star\sim 3-10 \times 10^9\, M_{\sun}$ wherein a maximum value $\approx 1.7$ is reached; however, $f_{\rm inf}\la 1$ is allowed at $1\sigma$ level.

For ETGs the resulting infall fraction reaches a maximum $f_{\rm inf}\approx 0.7$ around $M_{\star}\approx 3\times 10^{10}\, M_{\sun}$ and then it declines at larger masses, because of the combined decrease in efficiency and increase in metallicity, as shown in Fig. ~\ref{fig2_zfstar}. However, an infall fraction constant with mass $f_{\rm inf}\approx 0.3-0.4$ is allowed at $1\sigma$.

Interestingly, the fraction of the inflowing gas that is then removed by feedbacks $M_{\rm out}/M_{\rm inf}$ is always larger than $70\%$ for LTGs, while it is substantially lower for ETGs (cf. dashed lines in top panel of Fig.~\ref{fig4}). On the other hand, the fraction of inflowing mass eventually retained into stars $M_{\star}/M_{\rm inf}$ is larger for ETGs, reaching $60\%$ (see solid lines in top panel of Fig.~\ref{fig4}). This reflects the dilution needed for LTGs to keep the stellar metallicity low even in presence of a higher star-formation efficiency.

The bottom panel in Fig.~\ref{fig4} shows that only a small fraction of metals produced by stars are retained within the galaxy (i.e., in stars, ISM, and dust). We exclude the CGM from the budget since it does not enter in the galaxy mass and metal balances, though its composition carries some relevant information on complex inflow/outflow processes (see \citealt{Peeples2014}). For LTGs such a fraction is $\approx 20-30\%$ almost constant with stellar mass, as found by \cite{Peeples2014} for a large sample of spiral galaxies. Contrariwise, for ETGs we find that the fraction is increasing with stellar mass, reaching $\approx 60\%$.

\begin{figure}
\centering
\includegraphics[width=0.6\linewidth]{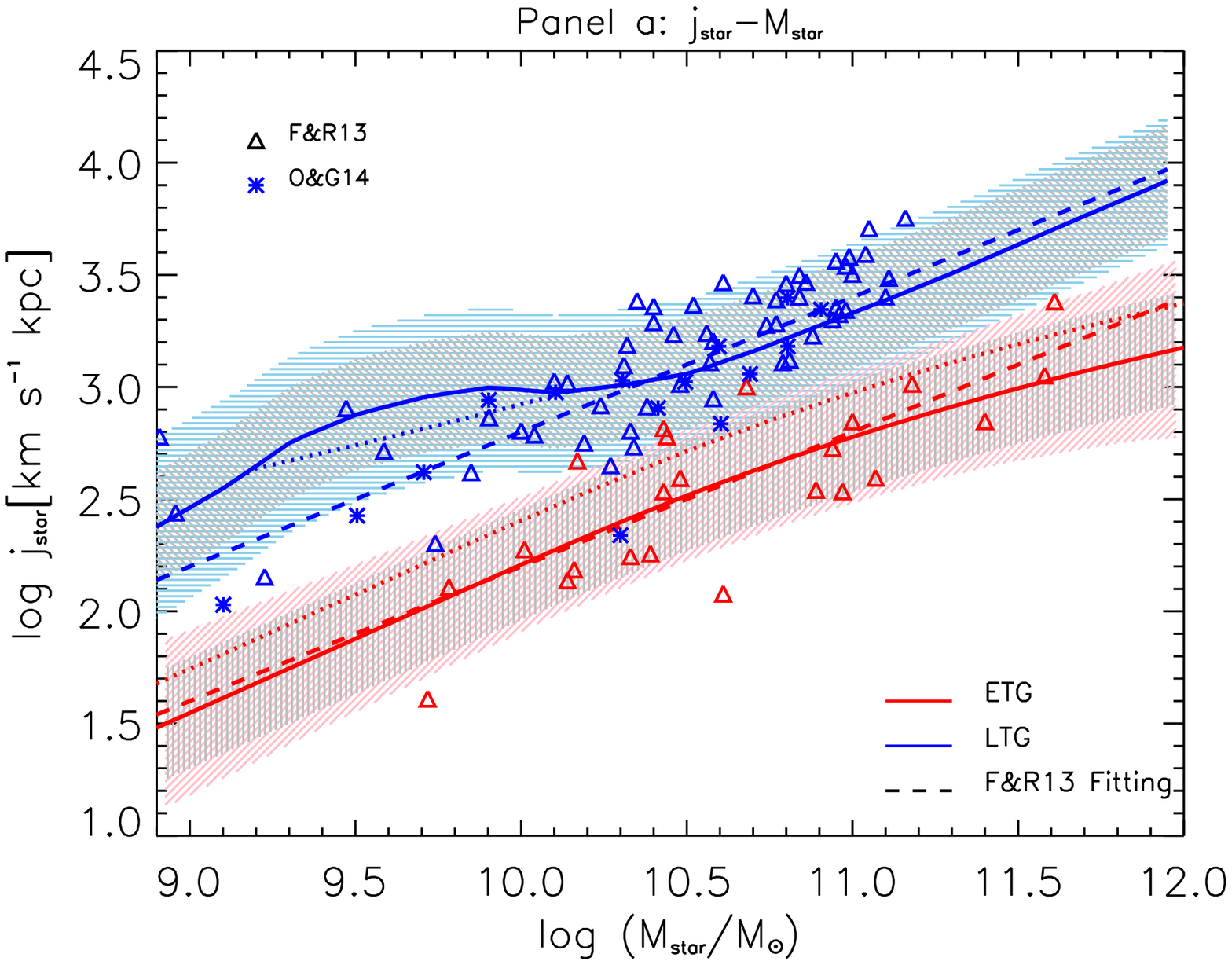}
\includegraphics[width=0.6\linewidth]{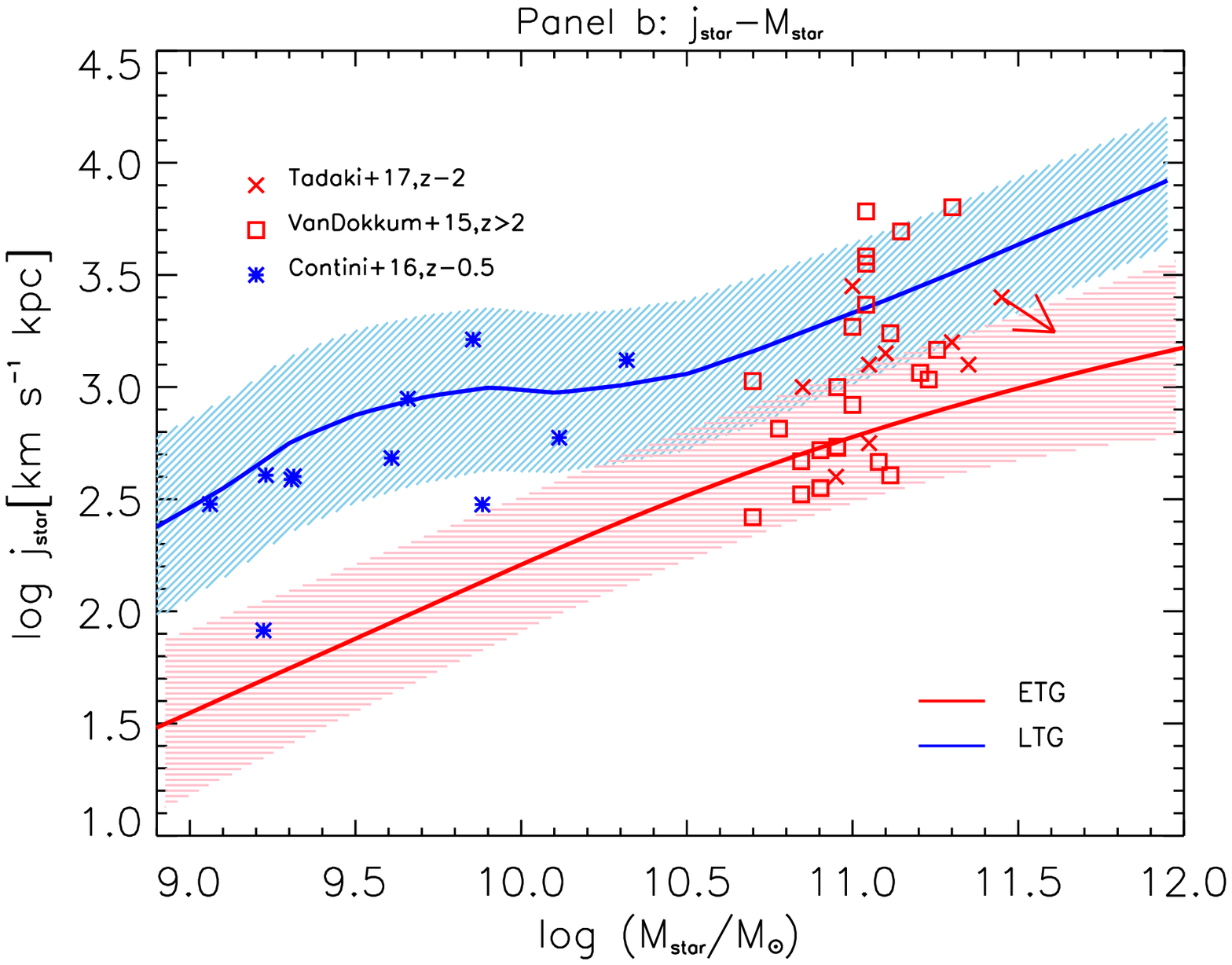}
\caption{Specific angular momentum $j_{\star}$ vs. the stellar mass $M_{\star}$ for ETGs (red lines) and LTGs (blue lines). Blue solid line is the result for LTGs with a retention fraction $f_{j}\approx 1$; dotted blue line applies when limiting the infall fraction $f_{\rm inf}\le 1$.
Red solid line is the result for ETGs taking into account stellar mass growth by dry mergers and a retention fraction $f_j\approx 0.64$; red dotted line refer to $f_j\approx 1$. In the top panel dashed lines represent the fitting formula $j_{\star}\propto M_{\star}^{0.6}$ adopted by \cite{Fall2013}. The colored shaded areas indicate the 1$\sigma$ uncertainty calculated by taking into account the variances of the parameters entering in Eq.~(\ref{eq_jinf}), while the grey shaded area includes only the intrinsic variance in the halo spin parameter $\lambda$ measured from numerical simulations. The blue and red triangles are data from \cite{Fall2013} for LTGs and ETGs, respectively. The blue stars are data for local spiral galaxies from \cite{Obreschkow2014}. In the bottom panel data for star-forming compact galaxies at $z\approx 2$ are reported: red squares are from \cite{vanDokkum2015} and red crosses from \cite{Tadaki2017};  data for disk galaxies at $z\approx 0.5$ from \cite{Contini2016} are also shown as blue stars. The red arrow shows explicitly the expected evolving direction of the high-$z$ ETG progenitors, after considering the growth in stellar mass envisaged by \cite{Belli2014} and $f_{j}= 0.64$.}
\label{fig5_JM}
\end{figure}

In Fig.~\ref{fig5_JM} (top panel) we illustrate relation between
specific angular momentum and stellar mass predicted after Eq.~(\ref{eq_jinf}) for LTGs and ETGs; this constitutes our main result. The differences in the inflowing fraction $f_{\rm inf}$, in the efficiencies $f_{\star}$, and in the formation redshift cooperate to yield distinct loci in the angular momentum vs. stellar mass plane for the two galaxy types (cf. Eq.~\ref{eq_jinf}).
To highlight the relevant dependencies, it is worth noticing that the handy approximation $f_{\rm inf}\approx y_Z\, f_{\star}/\langle Z_{\star}\rangle$ holds for both galaxy types. By plugging it in Eq.~(\ref{eq_jinf}), the specific angular momentum is seen to scale as
\begin{equation}\label{eq_jstar_approx}
j_{\star}\propto \lambda\, f_j\, f_{\star}^{-2/3+s}\, Z_{\star}^{-s}\, M_{\star}^{2/3}~;
\end{equation}
here $\lambda$ is independent of the host halo mass, and is assumed not to introduce additional dependence on the stellar mass. The inset of Fig.~\ref{fig3_finf} shows that the product $y_Z\,f_{\star}^{-2/3+s}\, Z_{\star}^{-s}\,M_\star^{0.15}\approx$ const. is different in normalization for each galaxy type but nearly independent of $M_\star$ for both (within the $1\sigma$ uncertainty). Since $s\sim 1$ the scaling $j_\star\propto f_\star^{1/3}$ applies, and hence the uncertainty in $f_\star$ only marginally contributes to that in $j_\star$.

Our result for LTGs (blue solid line) well describes the observed $j_{\star}$ vs. $M_\star$ relationship of discs. Note that we allow for $f_{\rm inf}\ga 1$, but we also plot (blue dotted line) the specific angular momentum under the condition $f_{\rm inf}\la 1$; as expected the estimates are within the respective $1\sigma$ uncertainty. Our result on $j_\star$ for LTGs implies a full retention of the initial specific angular momentum, i.e., $f_j\approx 1$. More quantitatively, a Montecarlo model fitting, that takes into account uncertainties in metallicities, $f_{\star}$ and $j_\star$, yields $f_j=1.11^{+0.75}_{-0.44}$. This is consistent within $1\sigma$ with the value around $0.8$ found by \cite{Fall2013}.

For ETGs the specific angular momentum (red dotted line) has been computed by using the efficiency at $z\approx 2$ and by assuming absence of evolution in the metal abundance (see Sect.~\ref{sec_fstar_et} and \ref{sec_z_et}). In addition, a shift in stellar mass by a factor $1.5$ has been applied to take into account mass additions by dry mergers at late times (red solid line; see Sect.~\ref{sec_fstar_et}). Comparison with local data for passive galaxies highlights that some room remains for a possible decrease of the specific angular momentum. Montecarlo model fitting, that takes into account uncertainties in $Z_\star$, $f_{\star}$ and $j_\star$ yields $f_j=0.64^{+0.20}_{-0.16}$. The average value may be explained by dry mergers at late times. For instance, if at later epochs the mass of the ETG progenitors is increased because of minor dry mergers with satellite galaxies (e.g., \citealt{Naab2009, Belli2014}), then a small decrease of the specific angular momentum can occur. The extent of this decrease is related to the sum of the initial momenta of the two companion galaxies and on their orbital momentum. For a limited mass increase of a factor $1.5$, a small decrease $j_{\star}\ga 1/1.5\approx 0.67\, j_{\rm inf}$ is expected, since the randomly oriented angular momenta of the companions partially cancels out (see also \citealt{Romanowsky2012}). Note that a value $f_j\la 0.1$, which would be needed to obtain the angular momentum of ETGs from the typical values for LTGs (and for the host halos), is excluded to more than $3\sigma$.

The colored shaded areas in Fig.~\ref{fig5_JM} represent the $1\sigma$ uncertainty in $j_{\star}$, which includes the uncertainties in $f_{\star}$ and metallicity, and the intrinsic variance in the exponent $s$ (see shaded areas in Fig.~\ref{fig3_finf}) and in the spin parameter $\lambda$; the variance in $\lambda$ actually dominates the overall scatter, as highlighted by the grey areas.

Focusing on the slope of the $j_\star-M_\star$ relation, \cite{Romanowsky2012} suggest that it can deviate from the expected value $2/3$, which stems from the definition $M_{\rm vir}=M_{\star}/f_{\star}\,f_b$ and from Eq.~(\ref{eq_jvir}), see also \cite{Catelan1996}. Our results in Fig.~\ref{fig5_JM} feature a running slope flatter than but close to $2/3$; specifically, by forcing a single power-law fitting, we get $j_{\star}\propto M_{\star}^{0.5}$ for LTGs and $j_{\star}\propto M_{\star}^{0.6}$ for ETGs. Interestingly, \cite{Fall2013} find a slope around $0.6$ for both, as indicated in Fig.~\ref{fig5_JM} (top panel) by the dashed lines.

To sum up, for LTGs the observed metallicity and star-formation efficiency imply that the fraction of the available baryons fueling star formation must be close to unity $f_{\rm inf}\approx 1$. Moreover, the specific angular momentum very well reproduces observations with a retention factor $f_j\approx 1$. On the other hand, for ETGs observations indicate that only a fraction $f_{\rm inf}\approx 0.4$ of the initial baryonic mass $f_b\, M_{\rm vir}$ must feed star formation; such a fraction of gas is endowed with low specific angular momentum, which turns out to be close to that observed for the stellar component in local passive galaxies. Data leave room for a small decrease $f_j\approx 0.64$ of specific angular momentum due to dry mergers possibly occurring between $z\la 1$ and the present time. Since for both galaxy types we find that the product $f_{\star}^{-2/3+s}\, Z_{\star}^{-s}$ only weakly depends on $M_{\star}$, then the slope of the $j_\star$ vs. $M_\star$ relationship is close to $2/3$, as observed for both galaxy types.

\section{Discussion}\label{sec_discussion}

In their thoughtful paper \cite{Romanowsky2012} reviewed the three most likely explanations for the observed location of ETGs and LTGs in $j_\star$  vs. $M_\star$ plane: (i) outflows of gas by some feedback mechanism or tidal stripping of the galactic halo; (ii) biased collapse plus merger scenario; (iii) pure merger driven evolution of LTGs into ETGs.

We have shown that current data on the star-formation efficiency and stellar metallicity naturally imply different infalling gas fractions for LTGs and ETGs, with average values $f_{\rm inf}\approx 1$ and $0.4$, respectively. These results strongly favor the biased collapse plus merger scenario, and they naturally locate ETGs and LTGs in two distinct loci of the $j_\star$  vs. $M_\star$ plane (cf. Fig.~\ref{fig5_JM}). While such a scenario is likely not the unique explanation for the observed $j_{\star}$ vs. $M_{\star}$ relationships in ETGs and LTGs, it points out the possibility that the history of star formation, hence $f_\star$ and $Z_\star$, knows about the assembly of the host DM halos and of their angular momentum.
Below we compare the predictions of the biased collapse plus mergers scenario to additional observational data and numerical simulations.

\subsection{The case of LTGs}\label{sec_dis_lt}

For LTGs we infer $f_{\rm inf}\approx 1$ and show that this value reproduces the observed $j_{\star}$ vs. $M_{\star}$ relationship, implying full retention of the specific angular momentum $f_j\approx 1$. Such results
are in line with the main assumption of the classical framework for disc formation, namely, that discs keep the overall specific angular momentum of their hosting halos (see \citealt{Fall1980, Mo1998, Mo2004, VanDenBosch2002}). A slow assembly of LTG discs is supported by the results of \cite{Hudson2015}, which showed that the ratio of the star to halo mass $M_\star/M_{\rm vir}$ keeps constant over a long cosmological timescale (from $z\approx 0.7$ to $0.3$); this is the epoch crucial for disc formation, as suggested by classical results on chemical and photometric evolution (see \citealt{Pezzulli2016}). Accurate spectrography for large samples of $z\approx 1$ star-forming galaxies shows that rotationally-dominated systems exhibit specific angular momentum lower by factors $1.5-2$ than local LTGs with the same stellar mass (see \citealt{Harrison2017, Swinbank2017}).
However, \cite{Contini2016} present evidence that LTGs at moderately low $z\sim 0.5$ fall on local $j_\star$ vs. $M_\star$ relationship within $1\sigma$; this possibly suggest rather weak dependence on the redshift such as $E(z)^{-1/6}$, cf. Eq.~(\ref{eq_jinf}); see also \cite{Burkert2016}.

The evolution of the angular momentum in galaxies has also been analyzed in \cite{Genel2015} by exploiting the results of the \textsl{Illustris} cosmological simulation (see \citealt{Vogelsberger2014a,Vogelsberger2014b, Genel2014a, DeFelippis2017}) and in \cite{Sokolowska2017} by using zoom-in simulation for MW like galaxies. These authors find that local LTGs retain $100\%$ of the specific angular momentum of their parent halos, likely enforced specific recipes for feedback and/or metal recycling. Their conclusion is also confirmed by the analyses of \cite{Zavala2016} and \cite{Lagos2017} based on the \textsl{EAGLE} numerical simulation (see also \citealt{Schaye2015}).

All in all, current observations and simulations indicate that feedback mechanisms (stellar winds, supernovae explosions and possibly AGNs) and ISM physics must cooperate to remove material from the galaxy star-forming regions, while cooling processes replace it with metal poor, high specific angular momentum gas; the overall outcome is that the metal content in star-forming regions is diluted and kept to low levels, while the specific angular momentum of the disc is increased.
All that occurs on cosmological timescales of order of many Gyrs (see \citealt{Molla2016}).

\subsection{The case of ETGs}\label{sec_dis_et}

\cite{Romanowsky2012} pointed out that the biased collapse scenario should be carefully considered in the case of ETGs, which apparently underwent angular momentum loss. Their main reservation toward biased collapse stems from a constraint for the normalization of the stellar specific angular momentum, which scales as $j_{\star} \propto f_j\, f_{\star}^{-2/3}$ under the assumption that the relation $j_{\rm vir}\propto M_{\rm vir}^{2/3}$ expected for DM halos (see \citealt{Catelan1996}) transfers to $j_{\star}\propto M_{\star}^{2/3}$ for the stellar component. As a consequence, the normalization of the  correlation $j_{\star}$ vs. $M_{\star}$ is constrained to be $f_j\, f_{\star}^{-2/3}\approx 0.5$ (cf. Eqs.~15 and 16 in \citealt{Romanowsky2012}). For ETGs they adopt the fitting formula of $f_{\star}$- $M_{\star}$ from \cite{Dutton2010}, obtaining $f_j\approx 0.1$.

On the other hand, we demonstrated that the chemistry and the star-formation efficiency of ETGs imply a small fraction of infalling gas mass $f_{\rm inf}\approx 0.4$. This parameter just quantifies the amount of biased collapse and it naturally decreases the normalization of $j_{\star}$ by a factor $\approx 2.5$ (since $j_\star\propto f_j\, f_\star^{-2/3}\, f_{\rm inf}^s$ with $s\approx 1$). We have shown that to reproduce observations a retention fraction $f_j\approx 0.64$ is needed; this can be accommodated for in terms of mass addition $\Delta M_\star/M_\star\la 0.5$ by late-time dry mergers.

One of the most relevant prediction of the biased collapse scenario is that the specific angular momentum has been imprinted in the ETG progenitors since the very beginning, with only minor changes related to later evolution in mass and size. This prediction can be tested by computing the angular momentum of the high-$z$ candidate progenitors of ETGs. Among the observed candidates there are $25$ compact star-forming galaxies at
$z\approx 2$ that have been studied in detail by \cite{vanDokkum2015}.
In particular, the observed structural and kinematical data of this optically selected sample allow to estimate the specific angular momentum of the galaxies, by exploiting the approximation of \cite{Romanowsky2012} $j\approx k_n\, V_{\rm rot}\, R_e$,  where $n$ is the \cite{Sersic1963} index. The median values for the sample are $n\approx 4$ ($k_4\approx 2.3$), $r_e\approx 1.4$ kpc, $V_{\rm rot}\approx 340$ km s$^{-1}$, yielding a median value  $j\approx 1000$ km s$^{-1}$ kpc, very close to that observed in local ETGs endowed with similar stellar mass $M_{\star}\approx 10^{11}\, M_{\sun}$.
More in detail, Fig.~\ref{fig5_JM} (bottom panel) illustrates that $18$ out of $25$ galaxies ($70\%$ of the sample) fall within $1\sigma$ from the $j_{\star}$ vs. $M_{\star}$ relationship of local ETGs.

\cite{Tadaki2017} presented estimates of the specific angular momentum for
$9$ optically selected star-forming galaxies at $z\approx 2$, observed with ALMA and detected at $870\,\mu$m. In Fig.~\ref{fig5_JM} (bottom panel) these galaxies are shown to exhibit a distribution in the $j_{\star}$ vs. $M_{\star}$ plane similar to that of the galaxies observed by \cite{vanDokkum2015}. These results suggest that most of such galaxies are in fact the progenitors of the local ETGs and that their specific angular momentum is imprinted at the epoch of formation with only minor subsequent changes, as predicted by our scenario. We stress the importance of analysing larger galaxy samples in order to further test this conclusion.

It is also interesting to compare these observational findings to the outcomes of recent numerical simulations, like \textsl{Illustris} (see \citealt{Vogelsberger2014a, Vogelsberger2014b, Genel2014a, Genel2015}). As for LTGs \cite{Genel2015} find in the simulation a $j_{\star}$ vs. $M_{\star}$ relation similar to the observed local one. For ETGs the situation is more complex. For a fraction of them, namely the galaxies with high final values of $j_{\star}$, the evolution is quite similar to that of LTGs. On the other hand, for simulate ETGs with low final values of $j_\star$, \cite{Genel2015} envisages two evolutionary paths: (i) a rapid initial growth in specific angular momentum combined with a later robust increase in mass by a factor of $\sim 10$ and roughly no change in specific angular momentum; (ii) a sudden drop of the specific angular momentum mainly imposed by a major merger. These authors also find that radio-mode feedback from AGNs helps in reducing the angular momentum, particularly for high mass galaxies. In fact, analyzing the \textsl{EAGLE} simulation (see \citealt{Schaye2015}), \cite{Lagos2017} put forward the possibility that even an early star formation followed by a rapid quenching can be effective in producing low angular momentum galaxies.

All in all, the analyses on simulated ETGs by \cite{Zavala2016} and \cite{Lagos2017} support a strong relation between the specific angular momentum of the stars and that of the DM in the inner star-forming region. Future data on specific angular momentum of massive high-$z$ galaxies will provide a crucial test for this scenario and a robust benchmark for next-generation numerical simulations of galaxy formation.

In the biased collapse scenario, the feedbacks (stellar and AGN) are key processes, since they partially offset cooling and regulate the fraction of inflowing gas. More specifically, in the case of ETGs, AGN feedback is required in order to stop the gas inflow. This yields a high stellar metallicity and a pronounced $\alpha$ enhancement (see \citealt{Matteucci1994, Romano2002, Thomas2005}) and keep the specific angular momentum low. Also the relationships between central black hole mass, stellar mass, and velocity dispersion can be explained in this context (see \citealt{Silk1998, Granato2001, Granato2004, DiMatteo2005, Lapi2014}). The impact of the biased collapse plus mergers scenario on the size evolution of galaxies at high redshift $z\ga 1$ will be discussed in a forthcoming paper.

\section{Summary and conclusions}\label{sec_sum}

We have investigated the origin, the shape, the scatter, and the cosmic evolution in the observed relationship between specific angular momentum $j_\star$ and the stellar mass $M_\star$ in ETGs and LTGs. Our main findings are summarized as follows.

\begin{itemize}

\item[1.] We have exploited the observed star-formation efficiency $f_\star$ and chemical abundance $Z_\star$ to infer the fraction $f_{\rm inf}$ of baryons that infall toward the central regions of galaxies (see Sect.~\ref{sec_fstar_z}); we find $f_{\rm inf}\approx 1$ for LTGs and $\approx 0.4$ for ETGs, weakly dependent on $M_{\star}$ (see Sect~\ref{sec_finf_jstar}) with an uncertainties of about $0.25$ dex.

\item[2.] We have highlighted that the infall fraction $f_{\rm inf}$ is the key variable in determining the distinct loci occupied by LTGs and ETGs in the $j_{\star}$ vs. $M_{\star}$ diagram, with ETGs featuring relatively lower specific angular momentum than LTGs, as observed (see Sect~\ref{sec_finf_jstar}).

\item[3.] We have estimated the fraction $f_j\equiv j_\star/j_{\rm inf}$ of the specific angular momentum associated to the infalling gas eventually retained in the stellar component; for LTGs we have found $f_j\approx 1.1^{+0.75}_{-0.44}$, which is consistent with the results from observations and simulations, and matches the standard disc formation picture (see Sect~\ref{sec_finf_jstar}). For ETGs we have found that $f_j\approx 0.64^{+0.2}_{-0.16}$, that can be explained by a late-time evolution due to dry mergers.

\item[4.] We have found that the dependencies of $f_{\star}$ and  $Z_{\star}$ on $M_{\star}$ conspire to make $j_\star\propto f_{\star}^{-2/3+s}\, Z_{\star}^{-s}$ weakly dependent on the stellar mass, with an overall shape close to $j_\star\propto M_\star^{2/3}$, see Sect~\ref{sec_finf_jstar}.

\item[5.] We have shown that the scatter in the observed $j_{\star}$ vs. $M_{\star}$ relationship for ETGs and LTGs mainly comes from the intrinsic variance in the halo spin parameter $\lambda$, while the uncertainties in star-formation efficiency $f_\star$ and stellar metallicity $Z_{\star}$ are minor contributors (Sect.~\ref{sec_finf_jstar}).

\item[6.] We have highlighted that the specific angular momentum $j_\star$ for most ($\sim 70\%$) of the observed star-forming galaxies at $z\sim 2$ is indeed very close to the local value for ETGs, as expected in our scenario (see Sect.~\ref{sec_dis_et}). Recent analyses of state-of-the-art numerical simulations (e.g., \citealt{Lagos2017}) start to find evidence that an early star formation quenching can imprint low specific angular momentum in the stellar component, in pleasing agreement with our scenario based on  biased collapse plus mergers.
\end{itemize}

All in all, we find that for LTGs the specific angular momentum steadily change over cosmological timescales following the external gas inflow, while for ETGs the specific angular momentum is mainly imprinted in a biased collapse at high-redshift, and then it possibly undergoes a minor decrease due to late-time dry mergers. Thus we argue the angular momentum of both galaxy types is mainly imprinted by nature (and in particular by the assembly history of their host DM halos) and not nurtured substantially by the environment.

\acknowledgements
We thank the anonymous referee for valuable suggestions that helped to improve our manuscript. We are grateful to E. Karukes and P. Salucci for stimulating discussions. Work partially supported by PRIN MIUR 2015 `Cosmology and Fundamental Physics: illuminating the Dark Universe with Euclid'. AL acknowledges the RADIOFOREGROUNDS grant (COMPET-05-2015, agreement number 687312) of the European Union Horizon 2020 research and innovation programme. HYW is supported by NSFC(11522324,11421303).

\clearpage

\appendix

\section{Specific angular momentum in spherical shells of halos}\label{sec_appendix1}

In this Appendix we use state-of-the-art, high-resolution $N-$body simulations to investigate the distribution of the specific angular momentum profile within dark matter halos, as a function of mass and redshift.

\subsection{Simulation and halo identification}

We exploit a $N-$body simulation based on the Gadget-2 code \citep{Springel2005}. The simulation adopted a flat $\Lambda$CDM cosmological model from WMAP9 constraints \citep{Hinshaw2013}, with $\Omega_{\Lambda}=0.718$, $\Omega_M=0.282$, $\Omega_b=0.046$, and $h=H_0/100$ km s$^{-1}$ Mpc$^{-1}=0.697$, $\sigma_{8}=0.817$ and $n_{s}=0.96$. The CDM density field is traced by $2048^{3}$ particles, each with mass $m_{\rm p}\approx 7.29\times 10^{7}\, M_{\sun}\, h^{-1}$,
from $z=120$ to $z=0$ in a cubic box of a side length $200$ Mpc h$^{-1}$. The gravitational force is softened isotropically on a comoving length scale of 2 $h^{-1}$ kpc (Plummer equivalent). We have $100$ snapshots from $z=20$ to $z=0$ equally spaced in the logarithm of the expansion factor.

The dark matter haloes are identified with FOF group algorithm \citep{Davis1985} and a linking length of $0.2\,b$, where $b$ is the mean interparticles separation. We resolve all groups with at least $20$ particles. Furthermore, we run SUBFIND \citep{Springel2001} to acquire the self-bound subhalo catalogue for each snapshot. We define the halo mass to be the mass contained in a spherical region (centred on the dominant subhalo particle with the minimum gravitational potential) with average density equals $200\, \rho_{\rm crit}$. In the calculation, we take the halo mass range $M_{\rm vir}\sim 10^{11}-10^{13}\, h^{-1}\, M_{\sun}$.

\subsection{Specific angular momentum profile}

\cite{Bullock2001} found a power-law approximation which describes the angular momentum reasonably well,
\begin{equation}\label{eq_bullockfit}
j_{z}(M)\propto M(<r)^{s}
\end{equation}
with $s$ roughly distributed over the halos like a Gaussian with average $s=1.3\pm 0.3$. Note that here $j_z(M)$ is the specific angular momentum projected to the direction of total angular momentum $J$.

\begin{deluxetable}{lcccccccccccccccc}
\tablewidth{0pt}
\tablecaption{Distribution of $s$ at different redshifts. \label{tab_sfit}}
\tablehead{\colhead{Parameter} & &
\colhead{$z=0$} & & \colhead{$z=1$} & & \colhead{$z=2$} & &
\colhead{$z=3$} & & \colhead{$z=4$}}
\startdata
$\mu$ & &1.120 & &1.043 & & 0.927 & & 0.817 & & 0.737\\
$\sigma$ & & 0.352 & & 0.364 & & 0.344 & & 0.320 & & 0.300\\
\enddata
\tablecomments{A Gaussian function with mean $\mu$ and variance $\sigma$ has been adopted.}
\end{deluxetable}

\begin{figure}
\centering
\includegraphics[width=0.45\linewidth]{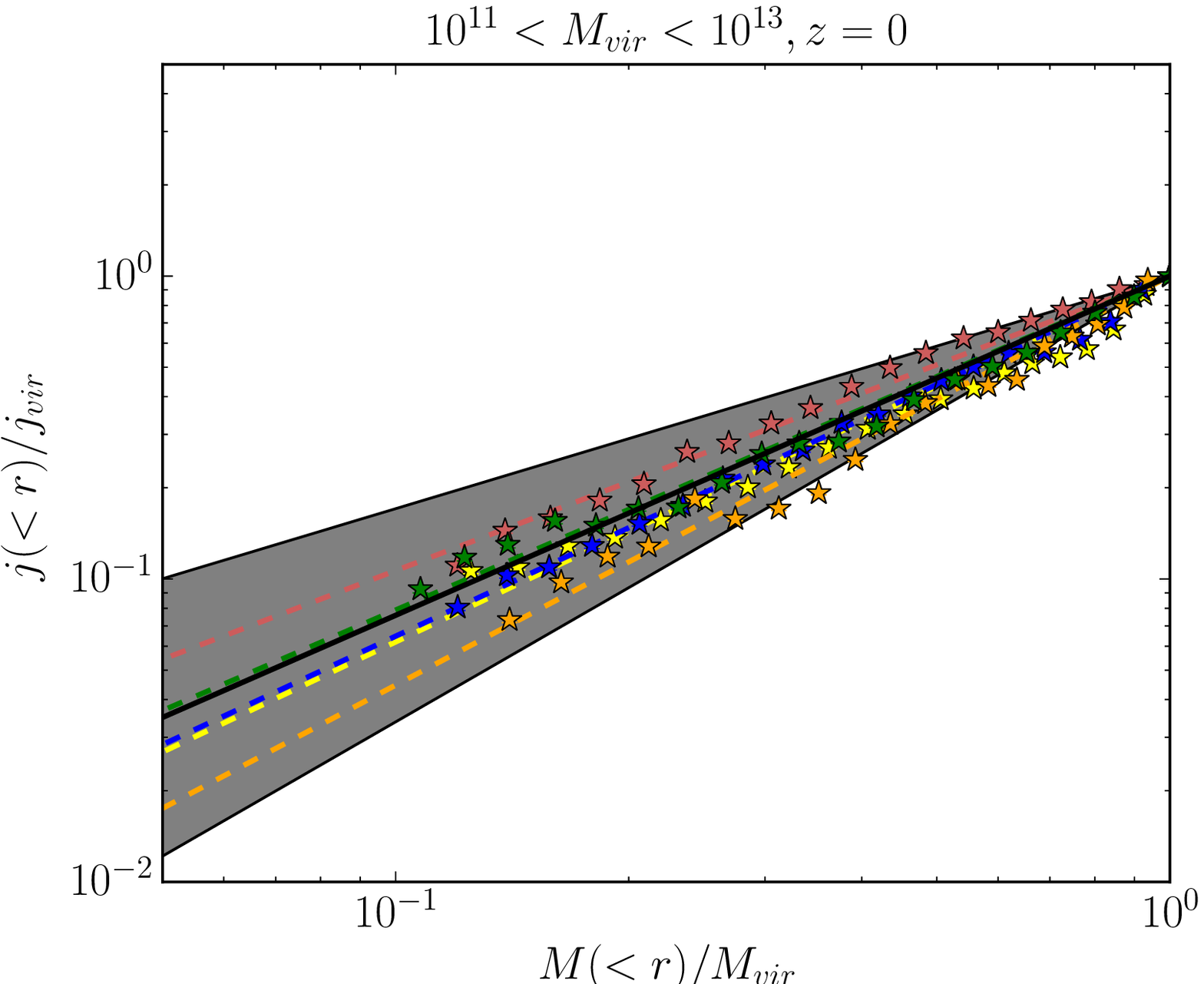}
\includegraphics[width=0.45\linewidth]{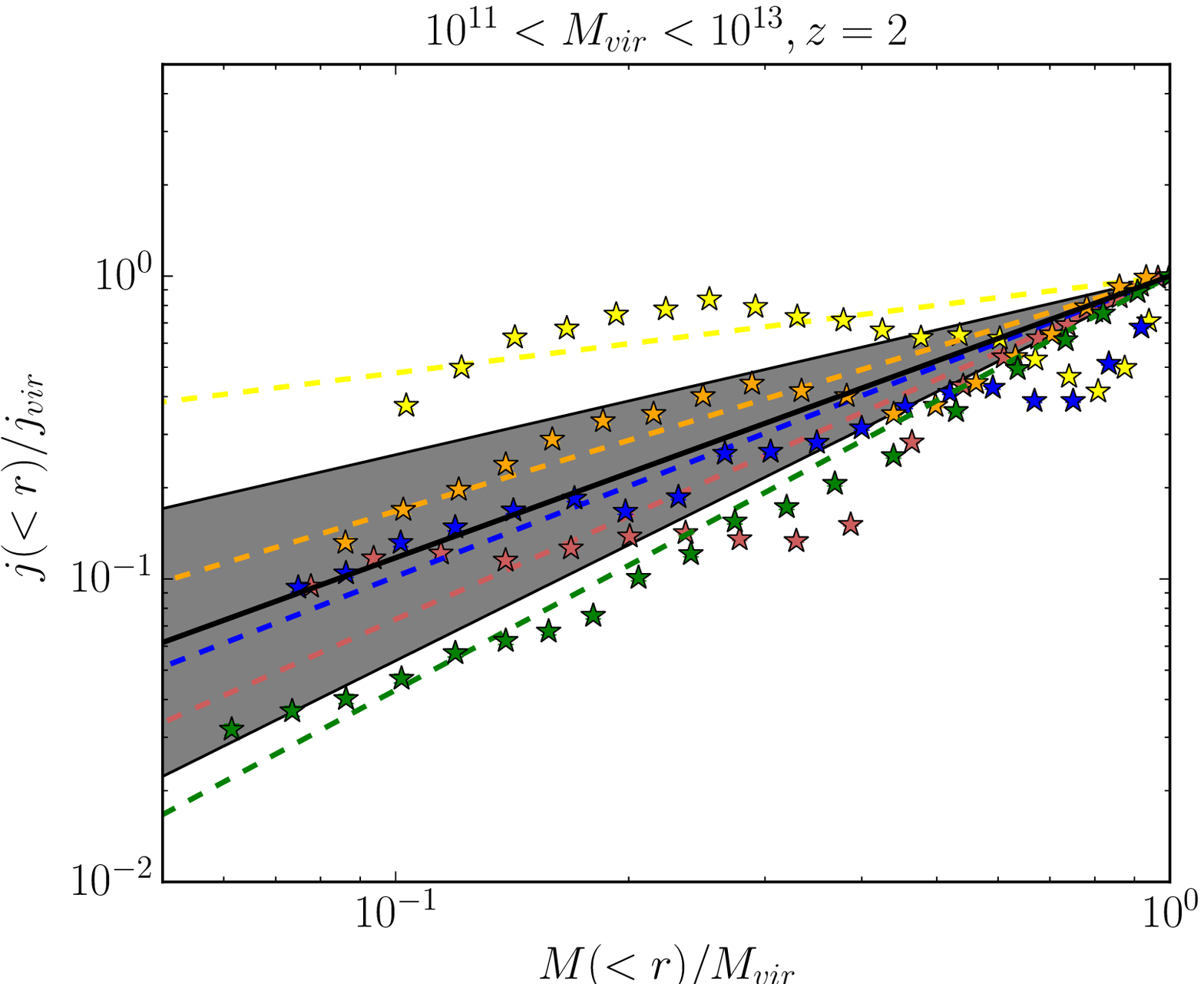}
\caption{Specific angular momentum vs. mass profile at redshifts $z=0$ and $z=2$. The lines with stars are the results for several randomly chosen halos in the sample, while the dashed lines are the fits with formula Eq.~(\ref{eq_improvedfit}). The black solid lines and the grey shaded areas show the mean profiles and the associated $1\sigma$ variance.}
\label{fig_JM_Prof}
\end{figure}

\begin{figure}
\centering
\includegraphics[width=0.45\linewidth]{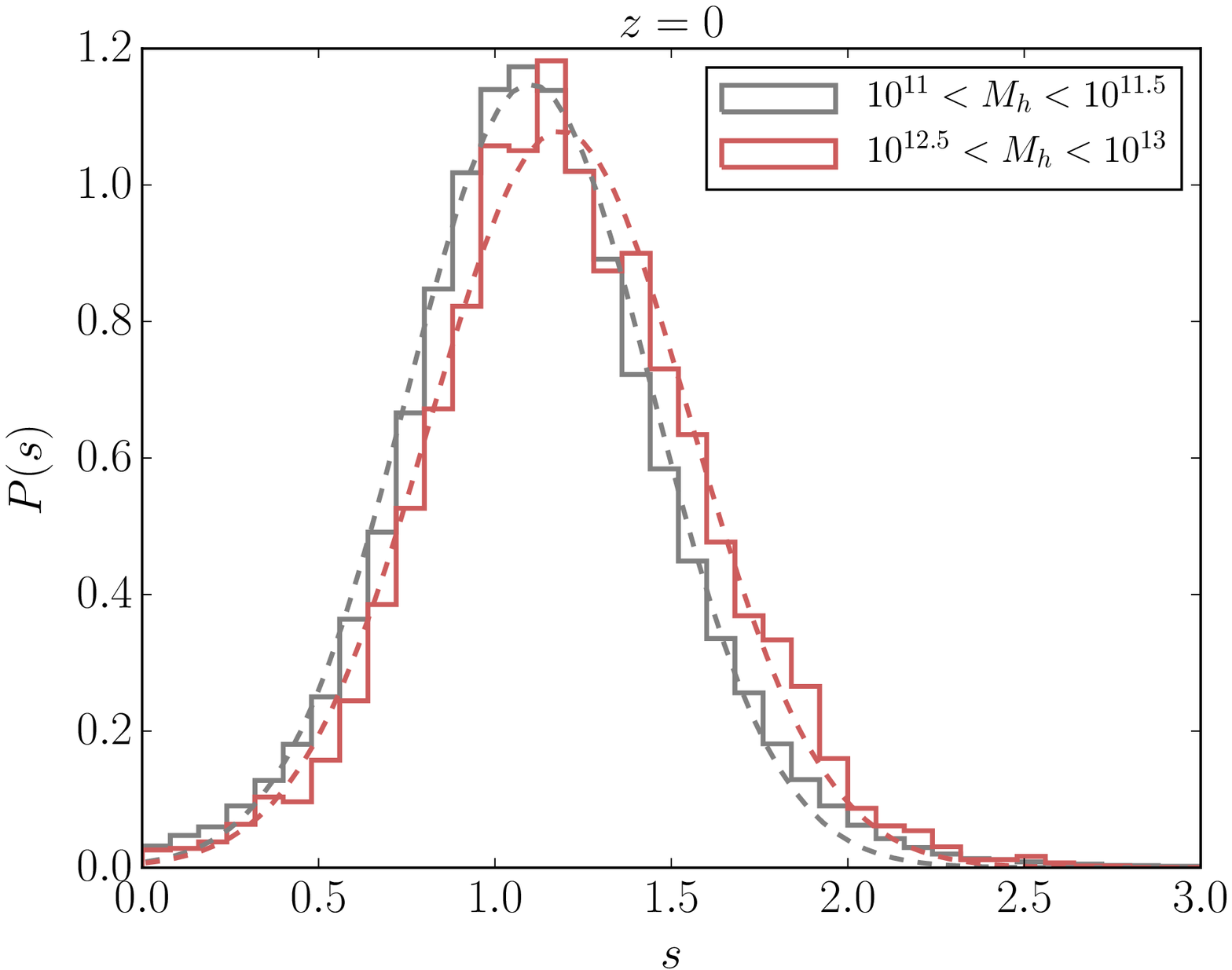}
\includegraphics[width=0.45\linewidth]{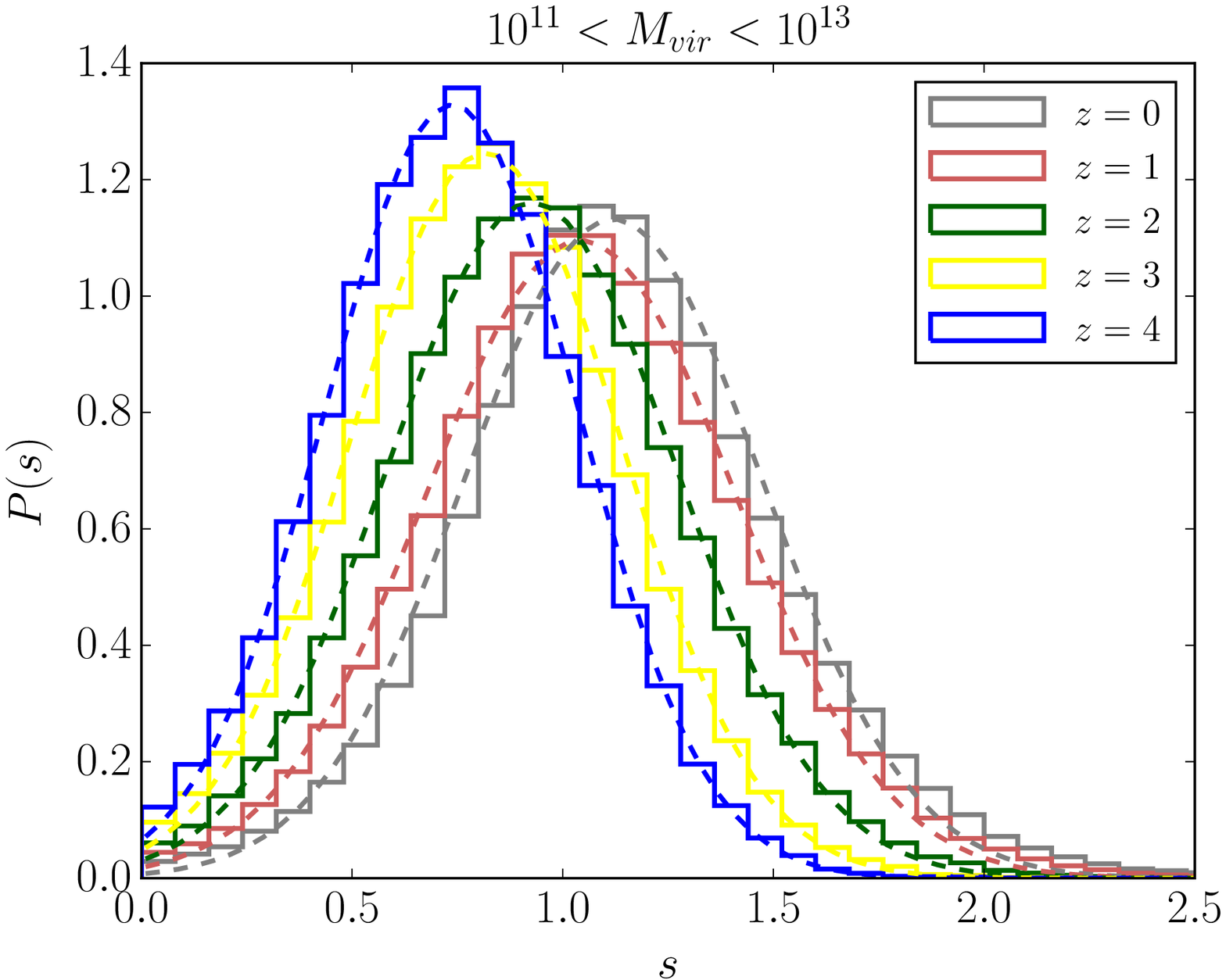}
\caption{Left panel: distribution of $s$ (power-law parameter in the specific angular momentum profile) for different halo masses at $z=0$. Right panel: distribution of $s$ at different redshifts for halo masses $M_{\rm vir}\sim 10^{11}-10^{13}\, h^{-1}\,M_{\sun}$. The dashed lines illustrate the gaussian fits, with the fitting parameters given in Table~\ref{tab_sfit}.}
\label{fig_sMhz}
\end{figure}

Here we look for a description of the relation between $j(M)$ and $M(<r)$, where $j(M)$ is the specific angular momentum (un-projected) within the shell with mass $M(<r)$. So we first divide each halo into shells between $0.1\, r_{\rm vir}$ and $r_{\rm vir}$. Then in each shell we calculate the specific angular momentum $j(<r)$ and mass $M(<r)$. Even though $j(<r)$ does not always increase monotonically with $M(<r)$, as shown by the data points in Fig.~\ref{fig_JM_Prof}, the power-law fitting does provide a useful rendition for the spherical distribution of $j$ on a statistical basis. Thus we use the formula
\begin{equation}\label{eq_improvedfit}
\frac{j(M)}{j_{vir}} = \left[\frac{M(<r)}{M_{vir}}\right]^{s}
\end{equation}
to fit our measurements in each halo of our samples. In addition, we checked the mass and redshift dependence of the power-law parameter $s$ in Fig.~\ref{fig_sMhz}. We find a very weak dependence on the mass and a decreasing $s$ with increasing $z$. The fitting parameters for $s$ with varying $z$ are listed in Table~\ref{tab_sfit}.

\section{Additional effects on the estimate of the infalling fraction}
\label{sec_appendix2}

In this Appendix we consider two additional effects that can alter somewhat the estimate of the infalling fraction $f_{\rm inf}$ discussed in Sect. 3.

The first effect concerns the metallicity of the infalling gas, that in Sect. 3 has been neglected. We now suppose that the gas mass $M_{\rm inf}$ infalling toward the central galaxy region is endowed with a metallicity $\langle Z_{\rm inf}\rangle$. Then the metal conservation Eq.~(10) must be modified into
\begin{equation}
y_Z\, M_\star + \langle Z_{\rm inf}\rangle\,M_{\rm inf} =  M_{Z,\rm gal} + M_{Z, \rm out}~,
\end{equation}
and along the same line of Sect. 3 we find that the infall fraction now reads
\begin{equation}
f_{\rm inf} = {f_\star\over 1-\langle Z_{\rm inf}\rangle/\zeta\,\langle Z_\star\rangle}\,\left({y_Z\over \zeta\,\langle Z_\star\rangle}-{M_{\rm Z,gal}\over \zeta\,\langle Z_\star\rangle\, M_\star}+{{M_{\rm gal}\over M_\star}}\right)~
\end{equation}
this replaces Eq. (11) of the main text, which is recovered when $\langle Z_{\rm inf}\rangle\ll \zeta\,\langle Z_\star\rangle$. The metallicity of the infalling gas is likely to be quite small $\langle Z_{\rm inf}\rangle\la$ a few $10^{-2}\, Z_\odot$, as suggested by various estimates for the intergalactic medium of local and high-redshift systems (for a review, see Madau \& Dickinson 2014); on considering that $\zeta\,\langle Z_\star\rangle\ga$ a few $10^{-1}\, Z_\odot$  (cf. Fig. 2), the correction to our estimate of $f_{\rm inf}$ is minor.

The second effect concerns the possibility that part of the outflowing gas falls back onto the galaxy, in the way of a galactic fountain circulation. We suppose that a fraction $\chi_{\rm rec}$ of the gas mass $M_{\rm out}$ outflown with metallicity $\langle Z_{\rm out}\rangle$ by feedback can fall back to the central galaxy after possible mixing with the metal poor gas in the outer regions.

The equation for the gas mass actually taking part in the galaxy formation process now writes
\begin{equation}
M_{\rm inf} =  M_{\rm gal} + (1-\chi_{\rm rec})\, M_{\rm out}
\end{equation}
and the metal mass conservation equation is modified into
\begin{equation}
y_Z\, M_\star + \langle Z_{\rm out}\,\rangle\,\chi_{\rm rec}\,M_{\rm out} =  M_{Z,\rm gal} + \langle Z_{\rm out}\rangle \, M_{\rm out}~.
\end{equation}
With respect to the main text equations, this amounts to a redefinition of the outflowing gas mass from $M_{\rm out}$
into $(1-\chi_{\rm rec})\, M_{\rm out}$. It is apparent that, in a one-zone model like that considered here, galactic fountain circulation does not affect the final value of $f_{\rm inf}$ which turns out to be unchanged with respect to Eq. (11). As a matter of fact, in detailed and spatially resolved chemical evolution approaches, galactic fountain is relevant in time delaying and spatially displacing metals (e.g., \citealt{Spitoni2013}).

\clearpage

\bibliography{shij_0602}

\end{document}